\documentclass[11pt]{article}
\usepackage{acl}

\definecolor{amethyst}{rgb}{0.6, 0.4, 0.8}

\usepackage{times}
\usepackage{latexsym}
\usepackage{microtype}
\usepackage{inconsolata}
\usepackage{graphicx}
\usepackage{booktabs}
\usepackage{multirow}
\usepackage{tabularx}
\usepackage{amsmath}
\usepackage{amssymb}
\usepackage{xcolor}
\usepackage{colortbl}
\usepackage{tikz}
\usepackage{booktabs}
\usepackage{multirow}
\usepackage{makecell}
\usepackage{graphicx}
\usetikzlibrary{arrows.meta,positioning,fit,calc}

\newcommand{\modelname}{ST-AudioLM}
\newcommand{\datasetname}{ST-AudioQA}
\newcommand{\encodername}{ST-Audio Encoder}

\newcommand{\map}{mAP}

\newcommand{\foa}{FOA}

\definecolor{oursblue}{HTML}{2F5BFF}
\definecolor{softgray}{HTML}{F3F5FA}
\definecolor{deepgray}{HTML}{2D3748}
\definecolor{auraltraceblue}{HTML}{EAF3FF}
\definecolor{revisioncolor}{HTML}{000000}
\newcommand{\revision}[1]{\textcolor{revisioncolor}{#1}}

\title{Spatio-Temporal Audio Language Modeling for Dynamic Sound Sources}

\def\authorBlock{
  \textbf{Oh Hyun-Bin${}^{1*}$\enspace
  Kazuki Shimada${}^{2}$\enspace
  Yuhta Takida${}^{2}$\enspace
  Kim Sung-Bin${}^{1}$\enspace
  Toshimitsu Uesaka${}^{2}$}\\
  \textbf{Takashi Shibuya${}^{2}$\enspace
  Kyeongyoon Lee${}^{4}$\enspace
  Tae-Hyun Oh${}^{5\dagger}$\enspace
  Yuki Mitsufuji${}^{2,3\dagger}$}\vspace{2mm}\\
  {\normalfont
  ${}^{1}$POSTECH\quad
  ${}^{2}$Sony AI\quad
  ${}^{3}$Sony Group Corporation\quad
  ${}^{4}$Sungkyunkwan University\quad
  ${}^{5}$KAIST}
}
\author{\authorBlock}

\begin{document}
\maketitle

\def\thefootnote{*}\footnotetext{Work done during an internship at Sony AI.}
\def\thefootnote{\textdagger}\footnotetext{Co-corresponding authors.}
\def\thefootnote{\arabic{footnote}}

\begin{abstract}

\revision{Sound events have semantic identities, locations, and trajectories, yet existing audio-language models typically encode clips as global event content.}
Conversely, sound event localization models track source directions over time but offer limited semantic coverage for language reasoning.
\revision{To bridge this gap, we introduce \datasetname{}, a spatio-temporal audio QA dataset and benchmark built from first-order ambisonic (\foa{}) renderings of static and moving sound sources.}
\revision{Each scene provides source identity, activity, direction, distance, and motion metadata, enabling dense trajectory supervision and QA about what is sounding, where it is, how it moves, and how sources relate.}
\revision{We further introduce \encodername{}, which jointly learns event semantics and time-resolved source trajectories, and \modelname{}, which exposes these audio tokens to an LLM for spatio-temporal audio QA.}
\revision{Experiments show that this representation improves the balance between semantic preservation and localization and yields stronger reasoning than static spatial and localization-oriented baselines.}
\revision{Beyond the controlled benchmark, \modelname{} shows zero-shot transfer to real recordings, with further gains after light adaptation.}
\revision{Project page: \url{https://st-audiolm.github.io/}.}

\end{abstract}

\section{Introduction}

Audio-language models have made rapid progress in recognizing, describing, and reasoning about sound events \citep{deshmukh2023pengi,gong2023ltu,tang2024salmonn,chu2023qwenaudio,kong2024audioflamingo}.
Yet most systems still treat an audio clip primarily as event labels or a caption to be generated.
This abstraction misses a basic property of everyday hearing: sound events unfold in space and time.
Humans use auditory location and motion to navigate, monitor danger, and decide how to act; a looming vehicle, receding footsteps, or speech fixed in place while another source passes by can change the interpretation of the same acoustic scene \citep{carlini2024auditory,neuhoff2001looming,seifritz2002looming}.
In such scenes, a sound is not merely a semantic category, such as \textit{siren} or \textit{speech}, but also an entity with a location, distance, and trajectory.
An audio-language model should therefore answer not only what is present, but also where each source is, whether it moves, how its distance changes, and how its motion relates to other sources.

\revision{Recent spatial audio-language work shows that large language models (LLMs) can use multichannel cues to reason about rendered sound scenes~\citep{bat2024,tang2024spatialaudio,wilkinghoff2026dspast,jiang2026sciphi,biswas2026owl,sakshi2025spur,zhu2026spatialomni}, while emerging benchmarks begin to consider dynamic audio~\citep{liu2025starbench,sridhar2026spatialaqa}.}
\revision{However, most models still focus on static or clip-level spatial attributes without explicitly exposing time-resolved source trajectories.}
Meanwhile, sound event localization and detection (SELD) models estimate time-varying directions and activities \citep{adavanne2018seldnet,shimada2022multiaccdoa,hu2025pseldnets}. 
\revision{Yet their constrained sound event taxonomies limit their use as broad-vocabulary semantic encoders for open-ended audio-language reasoning over AudioSet-scale event vocabularies~\citep{gemmeke2017audioset}.}
Together, these limitations reveal a gap between two lines of work:
audio-language models with broad semantics but weak temporal localization, and localization models with strong spatial inductive bias but limited language-level semantic coverage.

To bridge this gap, we contribute a dataset, an encoder, and an audio-language model.
First, we construct \datasetname, a controlled benchmark of static and moving sound sources rendered in first-order ambisonics (\foa{}).
We use \foa{} for its listener-centered full-sphere spatial coverage \citep{gerzon1973periphony,daniel2003spatial} and alignment with established SELD direction-estimation practice \citep{adavanne2018seldnet}, complementing prior binaural spatial audio-language work \citep{bat2024}.
Unlike event-only audio QA or static spatial QA, \datasetname{} is designed to evaluate binding between sound identity, time, and spatial state: it asks not only \textit{what} and \textit{where}, but also \textit{where at a given time}, \textit{how did it move}, and \textit{which source changed}.
The rendered spatio-temporal audio provides both dense trajectory supervision and structured source-level metadata for generating controlled QA, enabling training of both the spatio-temporal encoder and audio LLM.
Second, we introduce \encodername, a time-resolved spatio-temporal \foa{} audio encoder built on audio spectrogram transformers \citep{gong2021ast} that predicts AudioSet semantics together with activity, direction, and distance trajectories.
Third, we connect \encodername{} to a LLM~\cite{olmo20242} via audio tokens, yielding \modelname, and train it with a curriculum 
organized as A) single-source perception, B) two-source perception and grounding, and C) compositional source-time-space reasoning.

Experiments show that spatio-temporal audio QA requires more than recognizing events or localizing sources in isolation.
Encoders adapted from static spatial audio models~\cite{bat2024} struggle to track motion, while localization-oriented SELD models~\cite{hu2025pseldnets} lose event information needed for language questions.
\encodername{} addresses this tradeoff by learning representations that retain sound-event identity while tracking source direction and distance over time.
When used as audio tokens for an LLM, 
these representations outperform baselines based on temporalized static-encoder features or localization-oriented dynamic features across single-source perception, two-source grounding, and compositional motion questions.
\revision{Evaluations on real recordings demonstrate zero-shot transfer and gains from light adaptation~\citep{shimada2023starss23}; \modelname{} also leads the zero-shot comparison on a real-recording spatial-reasoning benchmark~\citep{liu2025starbench}.}
\revision{Together, these results suggest that spatio-temporal audio-language reasoning benefits from binding \textit{what is sounding} to \textit{where it is over time}.}
Our 
contributions are summarized as follows:
\begin{itemize}
    \item We propose \datasetname, a controlled spatio-temporal audio QA benchmark for static and moving sources, covering event recognition, time-anchored localization, motion, source grounding, and compositional reasoning.
    \item We introduce \encodername, a time-resolved \foa{} spatial audio encoder that preserves event semantics while predicting source direction and distance trajectories.
    \item We build \modelname, which equips an LLM with semantic and trajectory-aware \foa{} audio tokens, improving spatio-temporal perception, grounding, and compositional reasoning.
\end{itemize}

\section{Related Work}
\paragraph{Sound source localization models.}
Sound source localization and SELD studies provide a complementary line of work focused on estimating source activity and direction over time \citep{grumiaux2022survey,adavanne2018seldnet,cao2021improved}.
Formulations such as Multi-ACCDOA, which represents active sound events as Cartesian direction vectors, have been effective for localizing overlapping sources \citep{shimada2022multiaccdoa}.
Recent extensions further incorporate source distance estimation \citep{krause2024selddistance,diazguerra2024dcase} and exploit pretrained models, such as PSELDNets pretrained on large synthetic SELD data \citep{hu2025pseldnets}.
These models explicitly address the spatial and temporal localization problem that audio-language models often overlook.
However, their objectives are typically framed around localization and constrained sound event taxonomies.
This makes them effective for estimating source trajectories, but less suitable as general audio interfaces for LLMs, where broad semantic understanding and language-facing representations are essential.

\paragraph{Audio-language models for spatial reasoning.}
Audio-language models connect pretrained audio or speech encoders \citep{gong2021ast,elizalde2023clap,chen2023beats,radford2023whisper} to LLMs for captioning, question answering, and instruction following \citep{deshmukh2023pengi,gong2023ltu,tang2024salmonn,chu2023qwenaudio,kong2024audioflamingo}.
\revision{Recent systems further broaden general audio understanding through scaled open models, time-aware representations, and diffusion language backbones~\citep{chu2024qwen2,goel2025audioflamingo3,yang2026mossaudio,zhou2026diffa2}.}
Although these models improve language-level reasoning over everyday sounds using large-scale audio and audio-text supervision \citep{gemmeke2017audioset, kim2019audiocaps, drossos2020clotho, wu2023large}, they typically encode audio as global semantic content, leaving spatio-temporal source state implicit.
Recent spatial audio-language models begin to address this gap.
BAT introduces a spatial audio encoder for binaural audio and enables LLMs to answer controlled questions about sound source semantics and locations \citep{bat2024}.
Subsequent work extends auditory LLMs to multichannel spatial tasks such as far-field speech recognition \citep{tang2024spatialaudio}, improves spatial audio representations \citep{wilkinghoff2026dspast}, estimates source and room descriptors \citep{jiang2026sciphi}, or adds geometry-aware and plug-in spatial modules for spatial audio-language reasoning \citep{biswas2026owl,sakshi2025spur}.
\revision{Spatial-Omni~\citep{zhu2026spatialomni} integrates \foa{} encoding into a LLM for spatial audio understanding.
However, most spatial audio-language models still focus on static or clip-level spatial reasoning without explicit time-resolved source trajectories.}

\revision{Recent work has begun to address dynamic audio reasoning: STAR-Bench includes a broad real-recording evaluation~\citep{liu2025starbench}, while Sridhar et al. present dynamic spatial audio QA with rendered binaural scenes and BAT-style encoder adaptation~\citep{sridhar2026spatialaqa}.
In contrast, \datasetname{} provides controlled trajectory supervision and a hierarchy of source-state perception, grounding, and reasoning tasks, while \encodername{} produces dedicated semantic and trajectory tokens directly from \foa{}.}
Together with \modelname{}, this enables audio-language reasoning grounded in time-varying source states.

\section{\datasetname{}: Spatio-Temporal Audio QA Benchmark}
\label{sec:data}
\begin{table*}[t!]
\centering
\scriptsize
\setlength{\tabcolsep}{3pt}
\renewcommand{\arraystretch}{0.8}
\newcommand{\qacell}[4]{\textit{#1}\newline \textbf{Cases:} #2\newline \textbf{Q:} \texttt{#3}\newline \textbf{A:} \texttt{#4}}
\newcommand{\primset}[1]{\{#1\}}

\textbf{\revision{(a) Perception and grounding with primitive QA families}}
\vspace{0.25em}

\begin{tabularx}{\linewidth}{p{2.35cm} X X}
\toprule
\textbf{Primitive QA family} & \textbf{A: Single-source perception} & \textbf{B: Two-source perception and grounding} \\
\midrule
\textbf{Event semantics} &
\qacell{Detect or verify the event content of one source}{S, D}{What sound events are present?}{dog bark} &
\qacell{Detect or verify the event content of two sources}{SS, SD, DD}{What sound events are present?}{dog bark; clapping} \\
\addlinespace[0.25em]
\textbf{Movement status} &
\qacell{Decide whether one source moves}{S, D}{Is the sound moving or stationary?}{moving} &
\qacell{Decide whether a named source moves}{SS, SD, DD}{Is the clapping moving or stationary?}{moving} \\
\addlinespace[0.25em]
\textbf{Spatial location} &
\qacell{Localize one source over time}{S, D}{What is the azimuth of the sound at the end?}{back-right} &
\qacell{Localize a named source over time}{SS, SD, DD}{What is the distance of the clapping at the end?}{2.0--2.5 m} \\
\addlinespace[0.25em]
\textbf{Spatial change} &
\qacell{Estimate beginning-to-end change}{D}{Does the sound get closer or farther overall?}{closer} &
\qacell{Estimate change for a named source}{SD, DD}{Does the clapping move clockwise or counterclockwise?}{clockwise} \\
\addlinespace[0.25em]
\textbf{Source grounding} &
\qacell{Not instantiated for one-source clips}{--}{Not asked}{Not applicable} &
\qacell{Select the event matching a spatial or motion attribute}{SS, SD, DD}{Which sound moves clockwise, dog bark or clapping?}{clapping} \\
\bottomrule
\end{tabularx}

\vspace{0.65em}
\textbf{\revision{(b) Compositional source-time-space reasoning}}
\vspace{0.25em}

\begin{tabularx}{\linewidth}{p{4.0cm} p{6.0cm} X}
\toprule
\textbf{Compositional QA family} & \textbf{Composed primitive families} & \textbf{C: Compositional example} \\
\midrule
\textbf{Temporal relation change} &
\primset{Event semantics, Spatial location, Spatial change} &
\qacell{Ask whether a relation changes from beginning to end}{SD, DD}{Does dog bark start closer than clapping and end farther away?}{no} \\
\addlinespace[0.25em]
\textbf{Movement-conditioned spatial relation} &
\primset{Movement status, Spatial location} &
\qacell{Compare the moving and stationary sounds at a time anchor}{SD}{At the middle, is the moving sound higher than the stationary sound?}{no} \\
\addlinespace[0.25em]
\textbf{Cross-source trajectory relation} &
\primset{Event semantics, Spatial change} &
\qacell{Compare how the two source trajectories change}{SD, DD}{Does dog bark change distance more than clapping?}{yes} \\
\bottomrule
\end{tabularx}
\vspace{-2mm}
\caption{\textbf{\datasetname{} hierarchy.} \revision{(a)} Primitive QA families instantiated for single-source perception and two-source perception/grounding. \revision{(b)} Compositional reasoning integrates these primitives into source-time-space relation questions over two trajectories. S/D denote static/dynamic one-source clips; SS/SD/DD denote the three two-source motion cases. Controlled answers include event labels, yes/no, spatial bins, and movement labels.}
\label{tab:qa-overview-v2}
\end{table*}

Existing spatial audio QA benchmarks~\cite{bat2024, biswas2026owl} reduce spatial reasoning to static source locations, whereas real acoustic scenes require tracking sources as they move.
We therefore formulate spatial audio QA as a spatio-temporal grounding problem: each event should be tied to where it is over time, whether it moves, and how its direction and distance change.
\datasetname{} adopts the source trajectory, rather than a clip-level position, as the unit of supervision.

To build this benchmark, we render FOA audio and derive controlled short-answer QA pairs from structured source metadata.
For each sound source indexed by $k$, we define its state at time frame $t$ as
\begin{equation}\label{eq:1}
    s_k(t) = \{a_k(t), \mathbf{d}_k(t), r_k(t), y_k\},
\end{equation}
where $a_k(t)$ denotes binary source activity (active vs. inactive), $\mathbf{d}_k(t)$ is the listener-centered direction, $r_k(t)$ is metric distance, and $y_k$ is the 
audio label set associated with the clip.
Static sources are represented as zero-motion trajectories, 
giving a unified metadata schema for static and moving sources.
Sec.~\ref{sec:foa-scene-generation} describes how we generate spatio-temporal \foa{} audio, and Sec.~\ref{sec:qa-generation} describes how these source states are converted into QA pairs.
Additional details on audio rendering and QA generation are provided in Appendix~\ref{app:data-details}.

\subsection{Spatio-Temporal \foa{} Audio Generation}
\label{sec:foa-scene-generation}

\paragraph{Spatial audio simulation.}
We use SoundSpaces 2.0~\cite{chen2022soundspaces2} with Matterport3D~\cite{chang2017matterport3d} scene meshes to simulate room acoustics, following prior spatial audio-language work~\cite{bat2024}.
For each room, we sample listener and source anchors from navigable same-room candidates, then query SoundSpaces for \foa{} room impulse responses (RIRs) that encode source-to-listener direction, distance, and reflections.
This produces 10-second, 32 kHz \foa{} clips with controllable source trajectories and room-scale reverberation.

\paragraph{Static and dynamic sources.}
We represent both static and moving sources with a common 40-bin temporal schema.
Each 10-second clip is divided into 40 time bins, corresponding to one spatial state every 0.25 seconds.
For a moving source, the listener is fixed and the source follows a same-room trajectory between valid source anchors.
For a static source, the trajectory is constant, \ie, the same source position is used for all time bins.
To render motion, we query a per-bin \foa{} RIR and crossfade adjacent rendered segments, obtaining time-varying renderings of a sound source.
We retain only trajectories that pass audibility and RIR-stability checks; additional filtering and rendering details are given in Appendix~\ref{app:rendering-details}.

\paragraph{Sound events and mixtures.}
We use AudioSet clips~\citep{gemmeke2017audioset} as monaural event waveforms and spatialize each waveform by rendering it through the SoundSpaces \foa{} RIRs.
AudioSet provides the event content and label set, while SoundSpaces provides the spatial propagation and trajectory annotations.
A rendered scene contains either one source ($K=1$) or two sources ($K=2$).
Each source is rendered independently into \foa{} audio, and two-source scenes are mixed by summing the rendered \foa{} source signals channel-wise.
Single-source scenes contain either one static or one moving source; two-source scenes cover static--static (SS), static--dynamic (SD), and dynamic--dynamic (DD) motion configurations.
The resulting metadata provides time-varying source activity, direction, distance, and event labels.
These structured source states are used both for encoder supervision and for generating the QA pairs described next.

\subsection{Spatio-Temporal QA Generation}
\label{sec:qa-generation}

Using the structured source state $s_k(t)$, we instantiate controlled short-answer QA pairs from the rendered scene metadata.
The questions are designed to probe event identity, time-localized spatial state, movement status, spatial change, and relations between source trajectories.
Table~\ref{tab:qa-overview-v2} summarizes the hierarchy of question families and types.

\paragraph{Primitive QA families.}
We define five primitive QA families over the source state $s_k(t)$, corresponding to \revision{Table~\ref{tab:qa-overview-v2}(a)}:
\begin{itemize}[leftmargin=1.2em, itemsep=0em, topsep=0.2em, label={--}]
    \item \textit{Event semantics} covers event detection and verification, asking what sound events are present or whether a queried event is audible.
    \item \textit{Movement status} decides whether a source is moving or stationary.
    \item \textit{Spatial location} localizes a source at a beginning, middle, or end time anchor, using listener-centered azimuth, elevation, and distance bins.
    \item \textit{Spatial change} estimates how a source changes from the beginning to the end of the clip, such as clockwise versus counterclockwise, upward versus downward, or closer versus farther.
    \item \textit{Source grounding} selects the sound event that matches a given spatial or motion attribute in two-source scenes.
\end{itemize}

\paragraph{QA types.}
The QA types form a progression from perceiving a single source, to grounding sources in mixtures, to reasoning over relations between trajectories.
Type A, \textit{single-source perception}, asks whether the model can recover the event identity, location, movement status, and spatial change of one source trajectory, testing basic source-state understanding.
Type B, \textit{two-source perception and grounding}, places the same source-level cues in a mixture: the model must answer about a named event or select the event matching a spatial or motion attribute, thereby binding each event identity to the correct trajectory.
Type C, \textit{compositional source-time-space reasoning}, composes the primitive families into two-source relation questions, requiring the model to compare grounded source states across event identity, time, space, and motion (see \revision{Table~\ref{tab:qa-overview-v2}(b)}).

\section{Method}
\label{sec:method}

\begin{figure*}[t]
    \centering
    \includegraphics[width=0.9\linewidth]{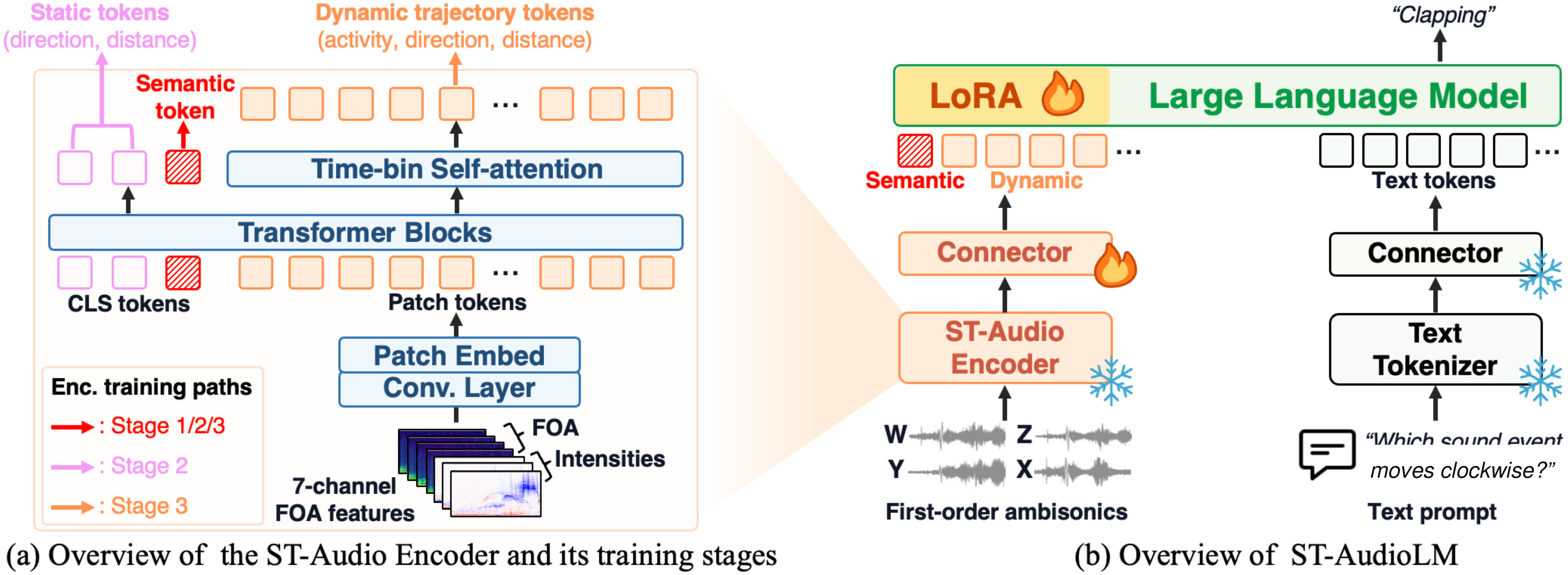}
    \vspace{-2mm}
    \caption{
    \textbf{Overview of \encodername{} and \modelname{}.}
    (a) \encodername{} converts \foa{} waveforms into seven-channel FOA features and uses an AST-style Transformer~\cite{gong2021ast} to produce semantic, static localization, and dynamic trajectory tokens, supervised by the corresponding objectives in three-stage encoder training. (b) For QA finetuning, the frozen \encodername{} provides one semantic token and dynamic trajectory tokens as audio tokens. A trainable two-layer MLP connector maps them into the LLM embedding space, and only the connector and LoRA adapters are trained on the staged spatio-temporal QA curriculum.}
    \label{fig:method-overview}
\end{figure*}

Figure~\ref{fig:method-overview} illustrates \modelname{}, which connects a time-resolved \foa{} audio encoder to an LLM for spatio-temporal audio QA.
We first describe \encodername{}, which learns event semantics together with source activity, direction, and distance trajectories (Sec.~\ref{sec:method-encoder}).
We then describe the audio-token interface and LLM training procedure used to turn the encoder representations into language-model inputs for perception, grounding, and compositional reasoning (Sec.~\ref{sec:method-llm}).

\subsection{\encodername{}: Spatio-Temporal FOA Audio Encoder}
\label{sec:method-encoder}

\paragraph{\foa{} frontend and static heads.}
\encodername{} uses an AST-style Transformer~\cite{gong2021ast} backbone over \foa{} spatial audio features.
Given a 10-second \foa{} waveform, we compute log-mel features from the four ambisonic channels $(W,X,Y,Z)$ and concatenate three acoustic intensity-vector features $(I_x,I_y,I_z)$, yielding a seven-channel time-frequency input, as commonly used for \foa{} SELD features~\citep{politis2020dataset}.
A $3{\times}3$ convolutional frontend fuses these channels before patch embedding, and the resulting patch sequence is processed by the Transformer.
To learn clip-level semantic and static spatial summaries, we prepend three learned global tokens: a semantic token, a direction-of-arrival (DoA) token, and a distance token.
The semantic token supports multi-label AudioSet prediction, while the DoA and distance tokens supervise static localization heads for clip-level DoA and source-listener distance.
This gives us a static \foa{} perception encoder, which we next turn into a time-resolved trajectory encoder.

\paragraph{Dynamic trajectory heads.}
The global tokens summarize clip-level semantics and static spatial state, but moving-source localization requires time-resolved evidence.
We therefore convert the final Transformer patch sequence into temporal trajectory representations.
Specifically, we reshape the final patch sequence into a grid $\mathbf{Z}\in\mathbb{R}^{T'\times F'\times C}$, where $T'$ and $F'$ are the temporal and frequency dimensions after patch embedding, and $C$ is the hidden dimension.
We then average over frequency to obtain a temporal representation:
\begin{equation}
    \mathbf{e}_{t} = \frac{1}{F'} \sum_{f=1}^{F'} \mathbf{Z}_{t,f}.
\end{equation}
The sequence $\{\mathbf{e}_t\}$ is linearly interpolated to 40 bins for each 10-second clip, corresponding to one representation every 250 ms.
Lightweight temporal self-attention layers contextualize these bin-level representations before prediction.
Per-bin dynamic heads then predict source activity, a 3D direction vector, and source-listener log-distance.
Static-source examples are represented as constant trajectories over time, allowing the same heads to train on both stationary and moving sources.

\paragraph{Encoder training.}
We train \encodername{} in three stages (see Fig.~\ref{fig:method-overview}(a)).
Stage 1 trains the semantic token for multi-label AudioSet prediction on static \foa{} renderings.
Stage 2 adds clip-level static localization by supervising the DoA and distance tokens, yielding a static \foa{} encoder with both event and spatial objectives.
Stage 3 converts this encoder into a time-resolved trajectory encoder: the dynamic heads are trained on moving-source trajectories, while static examples are replayed as constant trajectories over time.
To retain event recognition during this dynamic stage, we keep the semantic head active and distill semantic logits from \revision{a frozen Stage-1 semantic teacher}.

\subsection{\modelname{}: Spatio-Temporal Audio Reasoning with LLM}
\label{sec:method-llm}

\paragraph{Spatio-temporal audio token interface.}
After training \encodername{}, we remove all prediction heads and freeze the encoder.
For each audio clip, we use one semantic token and 40 temporal trajectory representations from the encoder as audio tokens, forming a 41-token audio sequence.
The semantic token provides a clip-level event representation, while the temporal trajectory representations preserve the temporal order of spatial information.
The global DoA and distance tokens are used only for encoder supervision and are not included in the LLM input.
A trainable two-layer MLP connector projects each audio token into the LLM embedding space.
Thus, the LLM receives both a semantic representation and a temporally ordered sequence of trajectory representations, rather than a single pooled audio vector (see Fig.~\ref{fig:method-overview}(b)).

\paragraph{LLM finetuning.}
During QA training, we freeze both \encodername{} and the base LLM weights, and update only the two-layer MLP connector and LoRA adapters.
We train these modules with a staged spatio-temporal QA curriculum, with training-pool sizes summarized in Table~\ref{tab:qa-curriculum-counts}.
Type A trains single-source perception, requiring the model to recognize event, location, and motion from the 41 audio tokens.
Type B adds two-source perception and grounding, where questions refer to named events or ask the model to select the source matching a spatial or motion attribute.
Type C adds compositional relation questions over grounded sources, combining event identity, time, location, motion, and source-to-source comparisons.

\begin{table}[t]
\centering
\small
\setlength{\tabcolsep}{5.5pt}
\renewcommand{\arraystretch}{0.8}
\begin{tabular}{c c c c}
\toprule
\textbf{Stage} & \textbf{Question Type} & \textbf{\# Tr. Samples} & \textbf{Percentages} \\
\midrule
I & A & 216K & 29.5\% \\
II & A, B & 516K & 70.3\% \\
III & A, B, C & 734K & 100\% \\
\bottomrule
\end{tabular}
\vspace{-2mm}
\caption{
\textbf{Staged QA curriculum for LLM finetuning.}
Training samples are cumulative over the listed question types: A is single-source perception, B adds two-source perception and grounding, and C adds compositional reasoning.
Percentages are relative to the Stage-III pool.
}
\label{tab:qa-curriculum-counts}
\end{table}
\vspace{-2mm}

\section{Experiments}
\label{sec:experiments}
We evaluate our approach by tracing the path from spatio-temporal audio representation to QA.
Encoder-level experiments examine whether \encodername{} captures both event identity and time-varying source state.
Audio-language experiments then assess \modelname{} on \datasetname{}, measuring single-source perception, two-source grounding, and compositional source-time-space reasoning.
Sec.~\ref{sec:exp-settings} describes the setup, metrics, and baselines; Sec.~\ref{sec:exp-results} presents the main analyses.

\subsection{Experimental Setup}
\label{sec:exp-settings}
\paragraph{Implementation details.}
Our encoder adapts the Spatial-AST backbone used in BAT~\citep{bat2024} to the \foa{} frontend in Sec.~\ref{sec:method-encoder}.
Stages 1--2 train this static \foa{} encoder on room-rendered AudioSet clips for multi-label event recognition and clip-level DoA/distance localization.
\revision{Stage 3 initializes \encodername{} from the Stage-2 checkpoint and learns time-resolved source trajectories through two training phases on single-source moving clips; Appendix~\ref{app:training-hyperparameters} provides stage-wise training details.}
For audio-language training, we use OLMo2-7B-Instruct~\citep{olmo20242}, freeze the audio encoder and base LLM, and train only the MLP connector and LoRA adapters.
\revision{We use LoRA rank 16 and $\alpha=32$; Type A uses connector and LoRA learning rates of $2\times10^{-4}$ and $10^{-4}$, while Types B and C use $10^{-4}$ and $5\times10^{-5}$, respectively.}

\paragraph{Encoder baselines.}
Table~\ref{tab:encoder-results} compares static and dynamic encoder baselines.
For static evaluation, \textit{Spatial-AST} is the binaural BAT encoder~\citep{bat2024}, and \textit{Spatial-AST-\foa{}} is our Stage-2 static \foa{} encoder before trajectory training.
For dynamic evaluation, \textit{Intensity-based DOA estimator} converts the \foa{} intensity vector into time-bin directions without learning.
\textit{Spatial-AST-\foa{} (temporal crop)} applies the trained static \foa{} encoder to short waveform crops centered at each trajectory time bin, predicting bin-level DoA and distance independently.
\textit{PSELDNets-mACCDOA + semantic/dist. heads} extends a dynamic PSELDNets architecture~\citep{hu2025pseldnets} to 355 AudioSet labels and source-listener distance prediction.

\paragraph{Audio-language baselines.}
\revision{Tables~\ref{tab:llm-results} and~\ref{tab:typec-results} compare \modelname{} with QA-trained baselines, and zero-shot evaluations of general audio-language models~\citep{chu2024qwen2,goel2025audioflamingo3,yang2026mossaudio,zhou2026diffa2} and the \foa{}-based Spatial-Omni~\citep{zhu2026spatialomni}.
Using each model's native or closest supported input (separate binaural channels for Qwen2-Audio, mono for the other general models, and \foa{} for Spatial-Omni), these comparisons test whether general audio-language or spatial-audio modeling suffices without trajectory-aware training.
\textit{BAT} uses the same QA pairs with corresponding binaural renderings, serving as a static binaural baseline.}
For controlled \foa{} comparisons, \textit{PSELDNets-mACCDOA + OLMo2} connects the dynamic PSELDNets-based encoder from Table~\ref{tab:encoder-results} to OLMo2, testing localization-oriented dynamic features for QA.
\textit{Spatial-AST-\foa{} + OLMo2} uses the frozen Stage-2 static \foa{} encoder, reshaping its patch tokens into 40 temporal tokens and appending the semantic token.
This matches \modelname{}'s 41-token interface but lacks \encodername{}'s trajectory-specific temporal self-attention, dynamic heads, and trajectory supervision.

\paragraph{Metrics.}
Encoder evaluation reports AudioSet \map{}, DoA MAE as 3D angular error in degrees, Dist. Acc over 0.5 m static distance bins, and Dist. MAE in meters for dynamic clips.
Traj.~Acc@20$^\circ$ counts a clip as correct only when all active time bins are localized within 20$^\circ$, measuring full-path rather than average localization.
For QA, event detection uses \map{}, event verification uses yes/no accuracy, and the other families use answer accuracy; Appendix~\ref{app:evaluation-details} details aggregation and parsing.

\begin{table}[t]
\centering
\footnotesize
\setlength{\tabcolsep}{2.4pt}
\renewcommand{\arraystretch}{1.2}

\textbf{(a) Static source evaluation}
\vspace{2pt}

\resizebox{\columnwidth}{!}{%
\begin{tabular}{l c c c c}
\toprule
\multirow{2}{*}{Model} & \multirow{2}{*}{Input} & \textbf{Semantic} & \multicolumn{2}{c}{\textbf{Spatial loc.}} \\
\cmidrule(lr){3-3}\cmidrule(lr){4-5}
& & \map{} $\uparrow$ & DoA MAE $\downarrow$ & Dist. Acc $\uparrow$ \\
\midrule
Spatial-AST~\cite{bat2024} & Binaural & \textbf{50.1} & \textbf{18.0}$^\circ$ & 67.2 \\
\rowcolor{auraltraceblue}
Spatial-AST-FOA & FOA & 47.7 & 19.2$^\circ$ & \textbf{74.1} \\
\bottomrule
\end{tabular}%
}

\vspace{5pt}
\textbf{(b) Dynamic source evaluation}
\vspace{2pt}

\resizebox{\columnwidth}{!}{%
\begin{tabular}{l c c c c c}
\toprule
\multirow{2}{*}{Model} & \multirow{2}{*}{Input} & \textbf{Semantic} & \multicolumn{2}{c}{\textbf{Spatial loc.}} & \textbf{Tracking} \\
\cmidrule(lr){3-3}\cmidrule(lr){4-5}\cmidrule(lr){6-6}
& & \map{} $\uparrow$ & DoA MAE $\downarrow$ & Dist. MAE $\downarrow$ & Traj. Acc@20$^\circ$ $\uparrow$ \\
\midrule
Intensity-based DOA estimator & FOA & -- & 23.9$^\circ$ & -- & 33.0 \\
Spatial-AST-FOA (temporal crop) & FOA & 45.2 & 41.4$^\circ$ & 0.64 m & -- \\
PSELDNets-mACCDOA + sem./dist. heads & FOA & 29.7 & \textbf{13.7}$^\circ$ & 0.38 m & 61.4 \\
\rowcolor{auraltraceblue}
\textbf{\encodername{} (ours)} & FOA & \textbf{62.8} & 13.8$^\circ$ & \textbf{0.32 m} & \textbf{62.3} \\
\bottomrule
\end{tabular}%
}
\vspace{-2mm}
\caption{\textbf{Encoder evaluation for static perception and dynamic tracking.}
(a) reports static event recognition, DoA estimation, and distance-bin prediction; the binaural Spatial-AST row is only a reference because \foa{} uses different renderings and input features.
(b) evaluates semantic prediction, DoA/distance errors, and whole-trajectory tracking on single-source moving clips.}

\label{tab:encoder-results}
\end{table}

\begin{table*}[t]
\centering
\scriptsize
\setlength{\tabcolsep}{1.7pt}
\renewcommand{\arraystretch}{1.0}
\resizebox{\linewidth}{!}{%
\begin{tabular}{l l c c c c c c c @{\hspace{6pt}} c c c c c c c c c c}
\toprule
\multirow{3}{*}{Type} &
\multirow{3}{*}{Model} &
\multicolumn{7}{c}{\textbf{A: Single-source perception}} &
\multicolumn{10}{c}{\textbf{B: Two-source perception and grounding}} \\
\cmidrule(lr){3-9}\cmidrule(lr){10-19}
& & \multicolumn{2}{c}{\textbf{Semantic}} & \multicolumn{2}{c}{\textbf{Spatial loc.}} & \multicolumn{2}{c}{\textbf{Spatial chg.}} & \textbf{Move}
& \multicolumn{2}{c}{\textbf{Semantic}} & \multicolumn{2}{c}{\textbf{Spatial loc.}} & \multicolumn{2}{c}{\textbf{Spatial chg.}} & \textbf{Move} & \multicolumn{3}{c}{\textbf{Source grounding}} \\
\cmidrule(lr){3-4}\cmidrule(lr){5-6}\cmidrule(lr){7-8}\cmidrule(lr){9-9}\cmidrule(lr){10-11}\cmidrule(lr){12-13}\cmidrule(lr){14-15}\cmidrule(lr){16-16}\cmidrule(lr){17-19}
& & \map{} & Y/N & DoA & Dist. & $\Delta$DoA & $\Delta$Dist. & Acc.
& \map{} & Y/N & Src. DoA & Src. Dist. & Src. $\Delta$DoA & Src. $\Delta$Dist. & Src. Move & Loc. & Chg. & Move \\
\midrule
-- & Random chance & 0.8 & 50.0 & 22.9 & 10.0 & 50.0 & 50.0 & 50.0
& 1.5 & 50.0 & 22.9 & 10.0 & 50.0 & 50.0 & 50.0 & 25.0 & 25.0 & 25.0 \\
\addlinespace[1pt]
\midrule
\multirow{5}{*}{Zero-shot} & Qwen2-Audio~\cite{chu2024qwen2} & 8.8 & 76.3 & 15.9 & 0.0 & 51.2 & 34.3 & 37.2
& 4.7 & 69.1 & 20.5 & 0.0 & 48.8 & 36.3 & 36.9 & 3.6 & 19.4 & 4.7 \\
& \revision{Audio Flamingo 3~\citep{goel2025audioflamingo3}} & \revision{17.8} & \revision{82.3} & \revision{19.3} & \revision{8.0} & \revision{35.2} & \revision{39.1} & \revision{43.1}
& \revision{7.8} & \revision{67.6} & \revision{18.4} & \revision{3.4} & \revision{37.2} & \revision{34.9} & \revision{40.8} & \revision{0.1} & \revision{0.4} & \revision{0.1} \\
& \revision{MOSS-Audio-8B~\citep{yang2026mossaudio}} & \revision{6.5} & \revision{79.0} & \revision{19.2} & \revision{21.0} & \revision{17.7} & \revision{48.6} & \revision{51.2}
& \revision{3.7} & \revision{62.1} & \revision{19.2} & \revision{19.5} & \revision{22.0} & \revision{49.2} & \revision{50.6} & \revision{13.4} & \revision{5.8} & \revision{1.2} \\
& \revision{DIFFA-2~\citep{zhou2026diffa2}} & \revision{3.8} & \revision{75.2} & \revision{17.8} & \revision{17.1} & \revision{32.1} & \revision{45.3} & \revision{51.1}
& \revision{2.8} & \revision{60.6} & \revision{18.2} & \revision{16.7} & \revision{32.6} & \revision{48.6} & \revision{49.1} & \revision{0.2} & \revision{0.1} & \revision{0.2} \\
& \revision{Spatial-Omni~\citep{zhu2026spatialomni}} & \revision{3.6} & \revision{68.6} & \revision{19.3} & \revision{8.6} & \revision{18.2} & \revision{34.0} & \revision{9.1}
& \revision{3.2} & \revision{44.4} & \revision{24.3} & \revision{8.1} & \revision{18.5} & \revision{29.9} & \revision{4.2} & \revision{8.1} & \revision{6.4} & \revision{8.6} \\
\addlinespace[1pt]
\midrule
\multirow{4}{*}{QA-trained} & BAT~\cite{bat2024} & 1.0 & 64.6 & 48.5 & 29.8 & 49.7 & 52.8 & 99.4
& 1.8 & 60.5 & 35.6 & 21.0 & 49.8 & 50.4 & 64.7 & 39.4 & 37.4 & 57.0 \\
& PSELDNets-mACCDOA + OLMo2 & 9.1 & 89.9 & 72.0 & 39.7 & 51.3 & 62.6 & 73.4
& 5.4 & 76.1 & 51.2 & 28.2 & 50.5 & 58.3 & 56.1 & 46.1 & 32.8 & 30.1 \\
& Spatial-AST-FOA + OLMo2 & 25.3 & 93.3 & 71.4 & 44.9 & 78.4 & 82.4 & 99.6
& 12.6 & 83.2 & 51.0 & 30.9 & 65.2 & 68.8 & \textbf{73.9} & 48.8 & 40.2 & \textbf{70.0} \\
\rowcolor{auraltraceblue}
& \textbf{\modelname{} (ours)} & \textbf{27.6} & \textbf{93.7} & \textbf{81.6} & \textbf{51.5} & \textbf{91.1} & \textbf{87.8} & \textbf{99.8}
& \textbf{14.3} & \textbf{83.8} & \textbf{54.3} & \textbf{32.8} & \textbf{71.6} & \textbf{70.0} & 66.2 & \textbf{49.7} & \textbf{47.4} & 59.3 \\
\bottomrule
\end{tabular}%
}
\vspace{-2mm}
\caption{\textbf{Audio-language QA results on \datasetname{}.}
All values are on a 0--100 scale; higher is better.
Type A evaluates single-source perception; Type B evaluates two-source perception and grounding.
Semantic columns report event-detection \map{} and event-verification accuracy; Spatial loc. reports DoA/distance accuracy, Spatial chg. reports beginning-to-end change accuracy, and Move reports moving/stationary accuracy.
For Type B, ``Src.'' denotes questions about the named sound event, and source grounding asks the model to select the event matching a spatial or motion attribute.
\revision{Zero-shot models use their native or closest supported input format. Qwen2-Audio and BAT use binaural audio rendered by SoundSpaces from the same scene and source metadata.}}
\label{tab:llm-results}
\end{table*}

\begin{table}[t]
\centering
\scriptsize
\setlength{\tabcolsep}{2.6pt}
\renewcommand{\arraystretch}{0.9}
\resizebox{\columnwidth}{!}{%
\begin{tabular}{l l c c c c}
\toprule
\multirow{2}{*}{Type} & \multirow{2}{*}{Model} &
\multicolumn{4}{c}{\textbf{C: Compositional source-time-space reasoning}} \\
\cmidrule(lr){3-6}
& & Temp. rel. & Move-spat. & Traj. rel. & Avg. \\
\midrule
-- & Random chance & 50.0 & 50.0 & 50.0 & 50.0 \\
\midrule
\multirow{5}{*}{Zero-shot} & Qwen2-Audio~\cite{chu2024qwen2} & 51.6 & 35.9 & 45.6 & 44.4 \\
& \revision{Audio Flamingo 3~\citep{goel2025audioflamingo3}} & \revision{49.8} & \revision{33.1} & \revision{49.8} & \revision{44.2} \\
& \revision{MOSS-Audio-8B~\citep{yang2026mossaudio}} & \revision{50.4} & \revision{33.1} & \revision{45.6} & \revision{43.0} \\
& \revision{DIFFA-2~\citep{zhou2026diffa2}} & \revision{49.7} & \revision{33.4} & \revision{45.1} & \revision{42.7} \\
& \revision{Spatial-Omni~\citep{zhu2026spatialomni}} & \revision{49.9} & \revision{28.2} & \revision{46.8} & \revision{41.6} \\
\midrule
\multirow{4}{*}{QA-trained} & BAT~\cite{bat2024} & 76.3 & 51.6 & 55.3 & 61.1 \\
& PSELDNets-mACCDOA + OLMo2 & \textbf{86.3} & 50.3 & 51.4 & 62.7 \\
& Spatial-AST-\foa{} + OLMo2 & 80.4 & 55.2 & 54.3 & 63.3 \\
\rowcolor{auraltraceblue}
& \textbf{\modelname{} (ours)} & 86.0 & \textbf{55.8} & \textbf{60.6} & \textbf{67.5} \\
\bottomrule
\end{tabular}%
}
\vspace{-2mm}
\caption{\textbf{Compositional QA results.}
All values are controlled-answer accuracy on a 0--100 scale; higher is better.
Temp. rel., Move-spat., and Traj. rel. denote temporal relation change, movement-conditioned spatial relation, and cross-source trajectory relation.
Avg. is computed over the reported Type-C families.}
\label{tab:typec-results}
\end{table}

\begin{table*}[t]
\centering
\scriptsize
\color{revisioncolor}
\begin{minipage}[t]{0.43\linewidth}
\centering
\textbf{(a) STARSS23 transfer and adaptation}\par\vspace{2pt}
\setlength{\tabcolsep}{3.0pt}
\renewcommand{\arraystretch}{1.0}
\begin{tabular}{l c c c c c}
\toprule
\multirow{2}{*}{Setting} & \multicolumn{1}{c}{Type A} & \multicolumn{1}{c}{Type B} & \multicolumn{3}{c}{Type C} \\
\cmidrule(lr){2-2}\cmidrule(lr){3-3}\cmidrule(lr){4-6}
& Overall & Overall & Move-spat. & Traj. rel. & Overall \\
\midrule
Zero-shot & 39.9 & 18.3 & 54.0 & 49.8 & 51.4 \\
Light adapt. & 59.7 & \textbf{59.4} & \textbf{56.4} & \textbf{67.8} & \textbf{63.5} \\
Real mixture & \textbf{67.1} & 59.1 & 44.2 & 62.6 & 55.7 \\
\bottomrule
\end{tabular}
\end{minipage}\hfill
\begin{minipage}[t]{0.55\linewidth}
\centering
\textbf{(b) STAR-Bench zero-shot evaluation}\par\vspace{2pt}
\setlength{\tabcolsep}{2.2pt}
\renewcommand{\arraystretch}{1.0}
\begin{tabular}{l l c c c c}
\toprule
Model & Input & Static loc. & Relation & Trajectory & Overall \\
\midrule
Random chance & -- & 33.3 & 33.3 & 33.3 & 33.3 \\
\midrule
Qwen2-Audio~\citep{chu2024qwen2} & 2ch & 42.2 & 22.2 & 13.3 & 37.9 \\
Audio Flamingo 3~\citep{goel2025audioflamingo3} & Mono & 36.6 & 52.8 & 26.7 & 37.4 \\
Spatial-Omni~\citep{zhu2026spatialomni} & \foa{} & 46.2 & 41.7 & 10.0 & 42.8 \\
\rowcolor{auraltraceblue}
\textbf{\modelname{} (ours)} & \textbf{\foa{}} & \textbf{47.9} & \textbf{63.9} & \textbf{60.0} & \textbf{50.4} \\
\bottomrule
\end{tabular}
\end{minipage}
\caption{\textbf{Real-recording evaluation.}
(a) STARSS23 transfer~\citep{shimada2023starss23}. Overall for each QA type is accuracy over all questions in its evaluation pool after answer normalization (chance for Type C: 50.0). \textit{Light adapt.} updates only the connector/LoRA after synthetic training; \textit{Real mixture} adds family-stratified STARSS23 QA amounting to approximately 5\% of the synthetic QA count at each curriculum stage (Appendix~\ref{app:starss23-results}).
(b) Zero-shot STAR-Bench evaluation~\citep{liu2025starbench} using each model's native or closest compatible input; none is trained on STAR-Bench. Overall reports micro-averaged accuracy. Values are percentages; higher is better.}
\label{tab:real-recording-results}
\end{table*}

\begin{table}[t]
\centering
\small
\color{revisioncolor}
\renewcommand{\arraystretch}{1.0}
\resizebox{1.0\linewidth}{!}{%
\begin{tabular}{l c c c c}
\toprule
\multirow{2}{*}{Setting} & \textbf{Semantic} & \multicolumn{2}{c}{\textbf{Spatial loc.}} & \textbf{Tracking} \\
\cmidrule(lr){2-2}\cmidrule(lr){3-4}\cmidrule(lr){5-5}
& \map{} $\uparrow$ & DoA MAE $\downarrow$ & Dist. MAE $\downarrow$ & Traj. Acc@20$^\circ$ $\uparrow$ \\
\midrule
\rowcolor{gray!10}\multicolumn{5}{l}{\textbf{(A) Semantic Distillation}} \\
Without & 51.0 & \textbf{12.2} & \textbf{0.30} & \textbf{73.0} \\
\rowcolor{auraltraceblue}
With (ours) & \textbf{62.8} & 13.8 & 0.32 & 62.3 \\
\midrule
\rowcolor{gray!10}\multicolumn{5}{l}{\textbf{(B) Temporal Resolution}} \\
10 bins & 58.5 & 14.1 & \textbf{0.31} & \textbf{65.4} \\
20 bins & 58.3 & 14.0 & \textbf{0.31} & 64.4 \\
\rowcolor{auraltraceblue}
40 bins (ours) & \textbf{62.8} & \textbf{13.8} & 0.32 & 62.3 \\
\midrule
\rowcolor{gray!10}\multicolumn{5}{l}{\textbf{(C) FOA Input Features}} \\
FOA log-mel only & 56.7 & 71.5 & 0.48 & 1.6 \\
$+$ IPD & 56.9 & 20.1 & 0.36 & 51.9 \\
\rowcolor{auraltraceblue}
$+$ Intensity vector (ours) & \textbf{62.8} & \textbf{13.8} & \textbf{0.32} & \textbf{62.3} \\
\bottomrule
\end{tabular}%
}
\caption{\textbf{Encoder design ablations.}
(A) semantic distillation, (B) temporal resolution, and (C) FOA input features. IPD denotes inter-channel phase difference. Variants share the Stage-3 training setup within each block. DoA/Dist. MAE are in degrees/meters; other metrics are percentages.}
\label{tab:encoder-ablations}
\end{table}

\subsection{Experimental Results and Analyses}
\label{sec:exp-results}
\revision{We analyze the results from encoder representation to audio-language reasoning, examining (i) semantic-preserving trajectory modeling, (ii) single- and two-source perception and grounding, (iii) compositional source-time-space reasoning, (iv) transfer to real recordings, and (v) key representation and training choices.
Appendix~\ref{app:starss23-transfer} provides detailed STARSS23 transfer results.}

\paragraph{Can an audio encoder model source trajectories while preserving event semantics?}
Table~\ref{tab:encoder-results} evaluates static spatial-semantic perception and dynamic moving-source tracking.
Table~\ref{tab:encoder-results}(a) includes binaural Spatial-AST only as a reference point because its input format and renderings differ from \textit{Spatial-AST-\foa{}}, while \textit{Spatial-AST-\foa{}} shows that our Stage-2 \foa{} encoder provides a strong static spatial-semantic representation before trajectory training.
In dynamic scenes, \textit{Spatial-AST-\foa{} (temporal crop)} shows that applying this static encoder over temporal windows is insufficient for trajectory tracking.
\textit{PSELDNets-mACCDOA + semantic/dist. heads} is a strong localization-oriented baseline, obtaining the best DoA error and competitive trajectory accuracy but substantially lower AudioSet \map{}.
In contrast, \encodername{} achieves the best semantic \map{} and distance error while remaining competitive with PSELDNets on dynamic localization.
This suggests that its main benefit is a joint representation that preserves broad event semantics while exposing time-resolved spatial cues for downstream language modeling.

\paragraph{Do trajectory-aware audio tokens improve single-source perception?}
Table~\ref{tab:llm-results} evaluates Type-A single-source perception, where the model must recognize the event and recover the source's spatial state over time.
\revision{Under zero-shot evaluation, general audio-language models retain some event-semantic ability but remain weak on localization, spatial change, and motion.}
Among QA-trained baselines, \textit{BAT} remains weak in event detection and spatial-change reasoning, while \textit{PSELDNets-mACCDOA + OLMo2} is limited by weak event semantics.
The strongest overall baseline, \textit{Spatial-AST-\foa{} + OLMo2}, uses the same 41-token interface as \modelname{} but derives its tokens from a static \foa{} encoder.
Compared with this, \modelname{} improves semantic recognition and spatial perception, with the clearest gains on trajectory-sensitive questions such as $\Delta$DoA.
This suggests that trajectory-supervised encoder representations make the audio tokens more informative for perceiving time-varying source states.

\paragraph{Do trajectory-aware tokens remain useful for grounding in mixtures?}
Type B in Table~\ref{tab:llm-results} moves from single-source perception to two-source mixtures, where the model must answer about a named event or select the event matching a spatial or motion attribute.
The static \foa{} baseline already performs strongly, showing that the 41-token interface provides a useful basis for mixture QA.
\modelname{} further improves semantic recognition, source-conditioned localization, and spatial-change grounding.
This suggests that trajectory-aware tokens remain useful beyond single-source perception, especially when spatial questions must be grounded to a queried sound event in a mixture.

\paragraph{How well do trajectory-aware tokens support compositional reasoning?}
Table~\ref{tab:typec-results} evaluates Type-C questions that combine event semantics, motion, spatial change, and cross-source comparison.
\revision{These zero-shot baselines remain below chance on average, showing that general audio-language modeling and \foa{} spatial modeling alone are insufficient for Type-C reasoning.}
\modelname{} achieves the best average accuracy, with the clearest gain on cross-source trajectory relations and performance comparable to the strongest localization-oriented baseline on temporal relation change.
However, movement-conditioned spatial relations remain only modestly above chance for all models, indicating that jointly resolving motion and spatial relations is still challenging.
Overall, these results suggest that trajectory-aware tokens help compositional source-time-space reasoning, while relation-level reasoning over grounded sources remains open.

\paragraph{\revision{Does source-time-space modeling transfer to real recordings?}}
\revision{Table~\ref{tab:real-recording-results}(a) evaluates transfer to STARSS23 recordings~\citep{shimada2023starss23}.
The zero-shot model retains semantic and some single-source perception abilities, while lightweight connector/LoRA adaptation substantially improves multi-source grounding and Type-C reasoning.
Mixing a small real-QA set into the synthetic curriculum yields the strongest Type-A transfer and improves Type-B grounding over zero-shot, although explicit adaptation remains strongest on Type-C.
Table~\ref{tab:real-recording-results}(b) further evaluates all models zero-shot on the real-recording subset of STAR-Bench~\citep{liu2025starbench}.
\modelname{} achieves the best result in every category and overall, with its clearest advantage on trajectory questions, providing evidence that its trajectory-aware representation transfers beyond synthetic rendering.}

\paragraph{\revision{What enables semantic-preserving spatio-temporal audio-language modeling?}}
\revision{Tables~\ref{tab:encoder-results}, \ref{tab:llm-results}, and~\ref{tab:typec-results} compare static and trajectory-trained representations at the encoder level and under the same LLM protocol, while Table~\ref{tab:encoder-ablations} isolates semantic distillation, temporal resolution, and FOA frontend features.
Distillation improves semantic preservation despite a localization trade-off, and the distilled encoder still exceeds PSELDNets in trajectory accuracy.
Among the tested resolutions, 40 bins give the best semantic \map{} and DoA error while keeping distance and trajectory tracking competitive.
Magnitude-only \foa{} severely degrades dynamic localization; IPD partially recovers it, but intensity-vector features perform best overall.
Appendix~\ref{app:curriculum-vs-joint} shows that the A$\rightarrow$B$\rightarrow$C curriculum progressively develops the three task types and better preserves Type-A/B than joint training under the same epoch budget.}

\section{Conclusion}

\revision{We present spatio-temporal audio-language modeling for dynamic sound sources.
To this end, we introduce \datasetname{}, a controlled, hierarchical source-time-space QA benchmark; \encodername{}, a time-resolved \foa{} encoder that preserves broad event semantics while modeling source trajectories; and \modelname{}, which exposes semantic and trajectory-aware audio tokens to an LLM.
Experiments show that these representations support dynamic localization and improve single-source perception, multi-source grounding, and compositional reasoning over static spatial and localization-oriented baselines.
\modelname{} also shows zero-shot transfer to real recordings, with further benefits from target-domain supervision.
Taken together, our results show that reasoning over moving sound sources requires more than static spatial understanding or localization alone; it calls for representations that bind broad event semantics to time-resolved source trajectories.}

\section{Limitations}
\revision{This work uses controlled synthetic rendering to obtain precise trajectory labels and systematic QA, but it cannot cover the full range of real acoustic conditions, microphone responses, source behaviors, noise, or outdoor environments.
We therefore assess transfer through zero-shot inference, light adaptation, and mixed real/synthetic QA training on STARSS23~\citep{shimada2023starss23}, together with zero-shot evaluation on STAR-Bench~\citep{liu2025starbench}.
Although these results demonstrate transfer beyond synthetic rendering, broader real-world evaluation remains necessary.}

The current benchmark also simplifies dynamic scenes.
\revision{Our dynamic rendering uses per-bin stationary \foa{} RIRs with crossfades. In our motion regime, the omitted Doppler shift is negligible; Appendix~\ref{app:rendering-details} quantifies this approximation, while high-speed pass-by events and continuously changing scattering paths remain outside our scope.}
Sources follow controlled 10-second trajectories, and many questions focus on beginning-to-end spatial change rather than arbitrary non-monotonic motion.
\revision{Movement-conditioned spatial-relation reasoning also remains an open challenge because it requires joint motion identification, source grounding, localization, and relation comparison.}
The main setting is limited to one- and two-source scenes, so denser mixtures, same-class source ambiguity, source entrances and exits, pauses, acceleration, and curved paths remain future directions.

\section{Ethical Considerations}
The system may inherit biases from AudioSet labels and the simulated scene distribution.
\revision{We do not annotate or infer speaker identity or other personal attributes, but speech-related and other audio labels may encode demographic or contextual bias.}
\revision{To support transparency and reproducibility, we will release the generation scripts and metadata filters.}
We avoid treating synthetic benchmark performance as a substitute for real-world robustness.

\section*{Acknowledgments}

This work was supported by the National Research Foundation of Korea (NRF) funded by the Korea government (MSIT) in 2026 (Grant No. RS-2024-00451947, Project Title: Interaction Framework for Connecting Users to Generative Experiences via Time-Space Expanded Multimedia: Centering on Social Communication Disorders (50\%); No. RS-2024-00358135, Corner Vision: Learning to Look Around the Corner through Multi-modal Signals (50\%)), the InnoCORE program of the Ministry of Science and ICT (26-InnoCORE-01), and the Artificial Intelligence Industrial Convergence Cluster Development Project funded by the Ministry of Science and ICT (MSIT, Korea) \& Gwangju Metropolitan City.

\makeatletter
\let\originalbibitem\bibitem
\newif\ifrevisionbibitem
\def\revisionbibkeys{gong2021ast,tang2024salmonn,chu2023qwenaudio,kong2024audioflamingo,bat2024,jiang2026sciphi,biswas2026owl,sakshi2025spur,wilkinghoff2026dspast,shimada2021accdoa,shimada2022multiaccdoa,diazguerra2024dcase,hu2025pseldnets,chang2017matterport3d,chen2022soundspaces2,olmo20242,shimada2023starss23,kim2019audiocaps,chu2024qwen2,dementyev2026phasecoder,goel2025audioflamingo3,yang2026mossaudio,zhou2026diffa2,liu2025starbench,zhu2026spatialomni}
\newcommand{\setrevisionbibcolor}[1]{%
  \revisionbibitemfalse
  \def\currentbibkey{#1}%
  \@for\revisionbibkey:=\revisionbibkeys\do{%
    \ifx\revisionbibkey\currentbibkey\revisionbibitemtrue\fi}%
  \ifrevisionbibitem\color{revisioncolor}\else\color{black}\fi
}
\def\bibitem{\@ifnextchar[\revisionbibitemwithlabel\revisionbibitemwithoutlabel}
\def\revisionbibitemwithlabel[#1]#2{%
  \color{black}\originalbibitem[#1]{#2}\setrevisionbibcolor{#2}}
\def\revisionbibitemwithoutlabel#1{%
  \color{black}\originalbibitem{#1}\setrevisionbibcolor{#1}}
\makeatother
\bibliography{bib/references}

@article{neuhoff2001looming,
  title = {An Adaptive Bias in the Perception of Looming Auditory Motion},
  author = {Neuhoff, John G.},
  journal = {Ecological Psychology},
  volume = {13},
  number = {2},
  pages = {87--110},
  year = {2001}
}

@article{seifritz2002looming,
  title = {Neural Processing of Auditory Looming in the Human Brain},
  author = {Seifritz, Erich and Neuhoff, John G. and Bilecen, Deniz and Scheffler, Klaus and Mustovic, Henrietta and Schachinger, Hartmut and Elefante, Raffaele and Di Salle, Francesco},
  journal = {Current Biology},
  volume = {12},
  number = {24},
  pages = {2147--2151},
  year = {2002}
}

@article{carlini2024auditory,
  title = {Auditory Localization: A Comprehensive Practical Review},
  author = {Carlini, Alessandro and Bordeau, Camille and Ambard, Maxime},
  journal = {Frontiers in Psychology},
  volume = {15},
  pages = {1408073},
  year = {2024}
}

@inproceedings{gemmeke2017audioset,
  title = {Audio Set: An Ontology and Human-Labeled Dataset for Audio Events},
  author = {Gemmeke, Jort F. and Ellis, Daniel P. W. and Freedman, Dylan and Jansen, Aren and Lawrence, Wade and Moore, R. Channing and Plakal, Manoj and Ritter, Marvin},
  booktitle = {ICASSP},
  year = {2017}
}

@inproceedings{gong2021ast,
  title = {{AST}: Audio Spectrogram Transformer},
  author = {Gong, Yuan and Chung, Yu-An and Glass, James},
  booktitle = {Interspeech},
  year = {2021}
}

@inproceedings{elizalde2023clap,
  title = {{CLAP}: Learning Audio Concepts From Natural Language Supervision},
  author = {Elizalde, Benjamin and Deshmukh, Soham and Al Ismail, Mahmoud and Wang, Huaming},
  booktitle = {ICASSP},
  year = {2023}
}

@inproceedings{chen2023beats,
  title = {{BEATs}: Audio Pre-Training with Acoustic Tokenizers},
  author = {Chen, Sanyuan and Wu, Yu and Wang, Chengyi and Liu, Shujie and Tompkins, Daniel and Chen, Zhuo and Che, Wanxiang and Yu, Xiangzhan and Wei, Furu},
  booktitle = {ICML},
  year = {2023}
}

@inproceedings{radford2023whisper,
  title = {Robust Speech Recognition via Large-Scale Weak Supervision},
  author = {Radford, Alec and Kim, Jong Wook and Xu, Tao and Brockman, Greg and McLeavey, Christine and Sutskever, Ilya},
  booktitle = {ICML},
  year = {2023}
}

@inproceedings{deshmukh2023pengi,
  title = {Pengi: An Audio Language Model for Audio Tasks},
  author = {Deshmukh, Soham and Elizalde, Benjamin and Singh, Rita and Wang, Huaming},
  booktitle = {NeurIPS},
  year = {2023}
}

@inproceedings{gong2023ltu,
  title = {Listen, Think, and Understand},
  author = {Gong, Yuan and Luo, Hongyin and Liu, Alexander H. and Karlinsky, Leonid and Glass, James},
  booktitle = {ICLR},
  year = {2024}
}

@inproceedings{tang2024salmonn,
  title = {{SALMONN}: Towards Generic Hearing Abilities for Large Language Models},
  author = {Tang, Changli and Yu, Wenyi and Sun, Guangzhi and Chen, Xianzhao and Tan, Tian and Li, Wei and Lu, Lu and Ma, Zejun and Zhang, Chao},
  booktitle = {ICLR},
  year = {2024}
}

@misc{chu2023qwenaudio,
  title = {{Qwen-Audio}: Advancing Universal Audio Understanding via Unified Large-Scale Audio-Language Models},
  author = {Chu, Yunfei and Xu, Jin and Zhou, Xiaohuan and Yang, Qian and Zhang, Shiliang and Yan, Zhijie and Zhou, Chang and Zhou, Jingren},
  year = {2023},
  eprint = {2311.07919},
  archivePrefix = {arXiv}
}

@inproceedings{tang2024spatialaudio,
  title = {Can Large Language Models Understand Spatial Audio?},
  author = {Tang, Changli and Yu, Wenyi and Sun, Guangzhi and Chen, Xianzhao and Tan, Tian and Li, Wei and Zhang, Jun and Lu, Lu and Ma, Zejun and Wang, Yuxuan and Zhang, Chao},
  booktitle = {Interspeech},
  year = {2024}
}

@inproceedings{kong2024audioflamingo,
  title = {{Audio Flamingo}: A Novel Audio Language Model with Few-Shot Learning and Dialogue Abilities},
  author = {Kong, Zhifeng and Goel, Arushi and Badlani, Rohan and Ping, Wei and Valle, Rafael and Catanzaro, Bryan},
  booktitle = {ICML},
  year = {2024}
}

@inproceedings{bat2024,
  title = {{BAT}: Learning to Reason about Spatial Sounds with Large Language Models},
  author = {Zheng, Zhisheng and Peng, Puyuan and Ma, Ziyang and Chen, Xie and Choi, Eunsol and Harwath, David},
  booktitle = {ICML},
  year = {2024}
}

@article{jiang2026sciphi,
  title = {{Sci-Phi}: A Large Language Model Spatial Audio Descriptor},
  author = {Jiang, Xilin and Braun, Sebastian and Gamper, Hannes},
  journal = {IEEE Open Journal of Signal Processing},
  year = {2026}
}

@inproceedings{biswas2026owl,
  title = {{OWL}: Geometry-Aware Spatial Reasoning for Audio Large Language Models},
  author = {Biswas, Subrata and Khan, Mohammad Nur Hossain and Islam, Bashima},
  booktitle = {ICLR},
  year = {2026}
}

@misc{sakshi2025spur,
  title = {{SPUR}: A Plug-and-Play Framework for Integrating Spatial Audio Understanding and Reasoning into Large Audio-Language Models},
  author = {Sakshi, S. and Lokegaonkar, Vaibhavi and Zhang, Neil and Duraiswami, Ramani and Ghosh, Sreyan and Manocha, Dinesh and Lu, Lie},
  year = {2025},
  eprint = {2511.06606},
  archivePrefix = {arXiv}
}

@misc{sridhar2026spatialaqa,
  title = {Spatial Audio Question Answering and Reasoning on Dynamic Source Movements},
  author = {Sridhar, Arvind Krishna and Guo, Yinyi and Visser, Erik},
  year = {2026},
  eprint = {2602.16334},
  archivePrefix = {arXiv}
}

@inproceedings{wilkinghoff2026dspast,
  title = {{DSpAST}: Disentangled Representations for Spatial Audio Reasoning with Large Language Models},
  author = {Wilkinghoff, Kevin and Tan, Zheng-Hua},
  booktitle = {ICASSP},
  year = {2026}
}

@article{grumiaux2022survey,
  title = {A Survey of Sound Source Localization with Deep Learning Methods},
  author = {Grumiaux, Pierre-Amaury and Kitic, Srdan and Girin, Laurent and Gu{\'e}rin, Alexandre},
  journal = {Journal of the Acoustical Society of America},
  year = {2022}
}

@article{adavanne2018seldnet,
  title = {Sound event localization and detection of overlapping sources using convolutional recurrent neural networks},
  author = {Adavanne, Sharath and Politis, Archontis and Nikunen, Joonas and Virtanen, Tuomas},
  journal = {IEEE Journal of Selected Topics in Signal Processing},
  year = {2018}
}

@inproceedings{shimada2021accdoa,
  title = {{ACCDOA}: Activity-coupled cartesian direction of arrival representation for sound event localization and detection},
  author = {Shimada, Kazuki and Koyama, Yuichiro and Takahashi, Naoya and Takahashi, Shusuke and Mitsufuji, Yuki},
  booktitle = {ICASSP},
  year = {2021}
}

@inproceedings{shimada2022multiaccdoa,
  title = {{Multi-ACCDOA}: Localizing and Detecting Overlapping Sounds from the Same Class with Auxiliary Duplicating Permutation Invariant Training},
  author = {Shimada, Kazuki and Koyama, Yuichiro and Takahashi, Shusuke and Takahashi, Naoya and Tsunoo, Emiru and Mitsufuji, Yuki},
  booktitle = {ICASSP},
  year = {2022}
}

@inproceedings{krause2024selddistance,
  title = {Sound Event Detection and Localization with Distance Estimation},
  author = {Krause, Daniel Aleksander and Politis, Archontis and Mesaros, Annamaria},
  booktitle = {EUSIPCO},
  year = {2024}
}

@inproceedings{diazguerra2024dcase,
  title = {Baseline Models and Evaluation of Sound Event Localization and Detection with Distance Estimation in {DCASE} 2024 Challenge},
  author = {Diaz-Guerra, David and Politis, Archontis and Sudarsanam, Parthasaarathy and Shimada, Kazuki and Krause, Daniel A. and Uchida, Kengo and Koyama, Yuichiro and Takahashi, Naoya and Takahashi, Shusuke and Shibuya, Takashi and Mitsufuji, Yuki and Virtanen, Tuomas},
  booktitle = {Workshop on Detection and Classification of Acoustic Scenes and Events},
  year = {2024}
}

@article{hu2025pseldnets,
  title = {{PSELDNets}: Pre-trained Neural Networks on a Large-scale Synthetic Dataset for Sound Event Localization and Detection},
  author = {Hu, Jinbo and Cao, Yin and Wu, Ming and Kang, Fang and Yang, Feiran and Wang, Wenwu and Plumbley, Mark D. and Yang, Jun},
  journal = {IEEE Transactions on Audio, Speech and Language Processing},
  year = {2025}
}

@inproceedings{chang2017matterport3d,
  title = {{Matterport3D}: Learning from {RGB-D} Data in Indoor Environments},
  author = {Chang, Angel and Dai, Angela and Funkhouser, Thomas and Halber, Maciej and Niessner, Matthias and Savva, Manolis and Song, Shuran and Zeng, Andy and Zhang, Yinda},
  booktitle = {International Conference on 3D Vision},
  year = {2017}
}

@inproceedings{chen2022soundspaces2,
  title = {{SoundSpaces} 2.0: A Simulation Platform for Visual-Acoustic Learning},
  author = {Chen, Changan and Schissler, Carl and Garg, Sanchit and Kobernik, Philip and Clegg, Alexander and Calamia, Paul and Batra, Dhruv and Robinson, Philip W. and Grauman, Kristen},
  booktitle = {NeurIPS Datasets and Benchmarks},
  year = {2022}
}

@inproceedings{politis2020dataset,
  title={A Dataset of Reverberant Spatial Sound Scenes with Moving Sources for Sound Event Localization and Detection},
  author={Politis, Archontis and Adavanne, Sharath and Virtanen, Tuomas},
  booktitle={Workshop on Detection and Classification of Acoustic Scenes and Events},
  year={2020}
}

@inproceedings{olmo20242,
  title = {{2 OLMo 2 Furious (COLM's Version)}},
  author = {Walsh, Evan Pete and Soldaini, Luca and Groeneveld, Dirk and Lo, Kyle and Arora, Shane and Bhagia, Akshita and Gu, Yuling and Huang, Shengyi and Jordan, Matt and Lambert, Nathan and Schwenk, Dustin and Tafjord, Oyvind and Anderson, Taira and Atkinson, David and Brahman, Faeze and Clark, Christopher and Dasigi, Pradeep and Dziri, Nouha and Ettinger, Allyson and Guerquin, Michal and Heineman, David and Ivison, Hamish and Koh, Pang Wei and Liu, Jiacheng and Malik, Saumya and Merrill, William and Miranda, Lester James Validad and Morrison, Jacob and Murray, Tyler and Nam, Crystal and Poznanski, Jake and Pyatkin, Valentina and Rangapur, Aman and Schmitz, Michael and Skjonsberg, Sam and Wadden, David and Wilhelm, Christopher and Wilson, Michael and Zettlemoyer, Luke and Farhadi, Ali and Smith, Noah A. and Hajishirzi, Hannaneh},
  booktitle = {COLM},
  year = {2025}
}

@inproceedings{shimada2023starss23,
  title={{STARSS23}: An audio-visual dataset of spatial recordings of real scenes with spatiotemporal annotations of sound events},
  author={Shimada, Kazuki and Politis, Archontis and Sudarsanam, Parthasaarathy and Krause, Daniel and Uchida, Kengo and Adavanne, Sharath and Hakala, Aapo and Koyama, Yuichiro and Takahashi, Naoya and Takahashi, Shusuke and Virtanen, Tuomas and Mitsufuji, Yuki},
  booktitle={NeurIPS Datasets and Benchmarks},
  year={2023}
}

@inproceedings{daniel2003spatial,
  title = {Spatial sound encoding including near field effect: Introducing distance coding filters and a viable, new ambisonic format},
  author = {Daniel, J{\'e}r{\^o}me},
  booktitle = {AES},
  year = {2003}
}

@article{gerzon1973periphony,
  title = {Periphony: With-height sound reproduction},
  author = {Gerzon, Michael A.},
  journal = {Journal of the Audio Engineering Society},
  year = {1973}
}

@inproceedings{kim2019audiocaps,
  title = {{AudioCaps}: Generating captions for audios in the wild},
  author = {Kim, Chris Dongjoo and Kim, Byeongchang and Lee, Hyunmin and Kim, Gunhee},
  booktitle = {NAACL-HLT},
  year = {2019}
}

@inproceedings{drossos2020clotho,
  title={Clotho: An audio captioning dataset},
  author={Drossos, Konstantinos and Lipping, Samuel and Virtanen, Tuomas},
  booktitle={ICASSP},
  year={2020}
}

@inproceedings{wu2023large,
  title={Large-scale contrastive language-audio pretraining with feature fusion and keyword-to-caption augmentation},
  author={Wu, Yusong and Chen, Ke and Zhang, Tianyu and Hui, Yuchen and Berg-Kirkpatrick, Taylor and Dubnov, Shlomo},
  booktitle={ICASSP},
  year={2023}
}

@misc{chu2024qwen2,
  title = {{Qwen2-Audio} Technical Report},
  author = {Chu, Yunfei and Xu, Jin and Yang, Qian and Wei, Haojie and Wei, Xipin and Guo, Zhifang and Leng, Yichong and Lv, Yuanjun and He, Jinzheng and Lin, Junyang and Zhou, Chang and Zhou, Jingren},
  year = {2024},
  eprint = {2407.10759},
  archivePrefix = {arXiv}
}

@misc{dementyev2026phasecoder,
  title = {{PhaseCoder}: Microphone Geometry-Agnostic Spatial Audio Understanding for Multimodal {LLM}s},
  author = {Dementyev, Artem and Zulfikar, Wazeer and Hersek, Sinan and Getreuer, Pascal and Kumar, Anurag and Kumar, Vivek},
  year = {2026},
  eprint = {2601.21124},
  archivePrefix = {arXiv}
}

@inproceedings{cao2021improved,
  title = {An improved event-independent network for polyphonic sound event localization and detection},
  author = {Cao, Yin and Iqbal, Turab and Kong, Qiuqiang and An, Fengyan and Wang, Wenwu and Plumbley, Mark D.},
  booktitle = {ICASSP},
  year={2021}
}

@inproceedings{goel2025audioflamingo3,
  title = {{Audio Flamingo 3}: Advancing Audio Intelligence with Fully Open Large Audio Language Models},
  author = {Ghosh, Sreyan and Goel, Arushi and Kim, Jaehyeon and Kumar, Sonal and Kong, Zhifeng and Lee, Sang-gil and Yang, Chao-Han Huck and Duraiswami, Ramani and Manocha, Dinesh and Valle, Rafael and Catanzaro, Bryan},
  booktitle = {NeurIPS},
  year = {2025}
}

@misc{yang2026mossaudio,
  title = {{MOSS-Audio} Technical Report},
  author = {Yang, Chen and Yu, Chufan and Chen, Hanfu and Zhu, Jie and Chen, Jingqi and Chen, Ke and Wang, Wenxuan and Wang, Yang and Jiang, Yaozhou and Jiang, Yi and Lin, Zhengyuan and Chen, Ziqi and Fei, Zhaoye and Liu, Chenghao and Yu, Donghua and Zhan, Jun and Yu, Kang and Huang, Kexin and Fan, Liwei and Chen, Mingshu and Cheng, Qinyuan and Li, Ruixiao and Li, Shimin and Wang, Songlin and Zhao, Xingjian and Gao, Yang and Gong, Yitian and Zhang, Yiyang and Xu, Zhe and Qiu, Xipeng},
  year = {2026},
  eprint = {2606.01802},
  archivePrefix = {arXiv}
}

@inproceedings{zhou2026diffa2,
  title = {{DIFFA-2}: A Practical Diffusion Large Language Model for General Audio Understanding},
  author = {Zhou, Jiaming and Cheng, Xuxin and Zhao, Shiwan and Jia, Yuhang and Liu, Cao and Zeng, Ke and Cai, Xunliang and Qin, Yong},
  booktitle = {Findings of ACL},
  year = {2026}
}

@inproceedings{liu2025starbench,
  title = {{STAR-Bench}: Probing Deep Spatio-Temporal Reasoning as Audio 4D Intelligence},
  author = {Liu, Zihan and Niu, Zhikang and Xiao, Qiuyang and Zheng, Zhisheng and Yuan, Ruoqi and Zang, Yuhang and Cao, Yuhang and Dong, Xiaoyi and Liang, Jianze and Chen, Xie and Sun, Leilei and Lin, Dahua and Wang, Jiaqi},
  booktitle = {ICLR},
  year = {2026}
}

@misc{zhu2026spatialomni,
  title = {{Spatial-Omni}: Spatial Audio Understanding Integration in Multimodal {LLM}s via {FOA} Encoding},
  author = {Zhu, Zhiyuan and Chen, Yixuan and Shao, Yiwen and Guo, Wenxiang and Pan, Changhao and Zhang, Yu and Wang, Yuxiang and Liu, Wei and Zhang, Houhua and Zeng, Chengkuan and Cheng, Wenbo and Liu, Yunxi and Yang, Rui and Yves, Steve and Bo, Liefeng and Zhao, Zhou},
  year = {2026},
  eprint = {2606.10738},
  archivePrefix = {arXiv}
}
\color{black}

\clearpage
\appendix
\hypersetup{linkcolor=black}
\section*{Contents}
\noindent\hyperref[app:model-training-details]{\textbf{A \ \ \ Model and Training Details}}\\
\hyperref[app:encoder-architecture]{\text{\quad A.1 \ \ \ \encodername{} Architecture}}\\
\hyperref[app:llm-connector]{\text{\quad A.2 \ \ \ Audio Tokens for the LLM}}\\
\hyperref[app:baseline-implementations]{\text{\quad A.3 \ \ \ Baseline Implementations}}\\
\hyperref[app:training-hyperparameters]{\text{\quad A.4 \ \ \ Training Configuration}}\\
\hyperref[app:evaluation-details]{\textbf{B \ \ \ Evaluation Details}}\\
\hyperref[app:curriculum-vs-joint]{\revision{\textbf{C \ \ \ Curriculum versus Joint Training}}}\\
\hyperref[app:real-data-transfer]{\textbf{D \ \ \ Real-Data Transfer}}\\
\hyperref[app:starss23-transfer]{\revision{\text{\quad D.1 \ \ \ STARSS23 Transfer QA Construction}}}\\
\hyperref[app:starss23-results]{\text{\quad D.2 \ \ \ STARSS23 Transfer and Adaptation}}\\
\hyperref[app:starbench-results]{\revision{\text{\quad D.3 \ \ \ STAR-Bench Zero-Shot Evaluation}}}\\
\hyperref[app:data-details]{\textbf{E \ \ \ Data Generation Details}}\\
\hyperref[app:rendering-details]{\revision{\text{\quad E.1 \ \ \ \foa{} Scene and Dynamic Audio Rendering}}}\\
\hyperref[app:qa-generation-details]{\text{\quad E.2 \ \ \ QA Generation Rules and Templates}}\\
\hyperref[app:dataset-statistics]{\text{\quad E.3 \ \ \ \datasetname{} Statistics}}\\
\noindent\rule{\linewidth}{0.2pt}
\hypersetup{linkcolor=darkblue}

\section{Model and Training Details}
\label{app:model-training-details}

We provide additional implementation details for \encodername{}, its interface with the language model, the baseline systems, and the training configurations used in our experiments.

\subsection{\encodername{} Architecture}
\label{app:encoder-architecture}

\paragraph{Input features.}
The encoder operates on a seven-channel time-frequency input.
The first four channels are log-mel features computed from the first-order ambisonic signal.
The remaining three channels are acoustic intensity-vector features derived from the same \foa{} signal.
We stack the mel and intensity features along the channel dimension and pass them through a $3{\times}3$ convolutional frontend before AST-style patch embedding.
This frontend fuses spectral, ambisonic, and intensity cues before transformer processing.
\revision{We sample the waveforms at 32 kHz and compute a 1,024-point STFT using a 1,024-sample Hann window and a 320-sample hop.
The resulting power spectra are projected onto 128 mel bands spanning 50 Hz to 14 kHz and compressed to decibels with an 80 dB range.
For $q\in\{X,Y,Z\}$, the intensity feature is the mel projection of
$\operatorname{Re}(W^{*}q)/\bigl(|W|^2+(|X|^2+|Y|^2+|Z|^2)/3+\epsilon\bigr)$.}

\paragraph{\revision{Backbone configuration.}}
\revision{The AST backbone uses 768-dimensional embeddings, 12 Transformer blocks, 12 attention heads, an MLP ratio of 4, and $16{\times}16$ patches with stride 16.
The frontend feature map is resized to 1,024 time frames before patch embedding.}

\paragraph{Static spatial pretraining.}
We first train the \foa{} encoder with a Spatial-AST-style objective~\citep{bat2024}.
This stage uses global representations to predict AudioSet event semantics, and static direction and distance.
The semantic representation is trained with AudioSet multi-label supervision, while the direction and distance representations are trained with static localization losses.
This provides a spatial-semantic initialization before introducing moving-source trajectory supervision.

\paragraph{Dynamic trajectory modeling.}
For dynamic trajectory modeling, we reuse the transformer patch sequence from the pretrained encoder instead of adding a separate audio frontend.
The post-transformer patch sequence is reshaped into a time-frequency grid, pooled over frequency, and represented as 40 temporally ordered states, matching the 40 spatial states used in the renderer.
We apply a self-attention temporal modeling block over these states.
The dynamic heads predict source activity, 3D direction, and log-distance for each time bin.
During dynamic training, we retain the global semantic representation and use semantic supervision and distillation to reduce degradation in event recognition.
\revision{The temporal module consists of two self-attention blocks with eight heads and an MLP ratio of 4.
Its prediction heads output one activity logit, a normalized 3D direction vector, and scalar log-distance per time bin.}

\paragraph{Static examples as constant trajectories.}
Static clips are represented as constant trajectories.
The same direction and distance target is repeated across all 40 time bins, so a stationary source becomes a zero-motion special case of the dynamic formulation.
This allows the temporal heads to learn moving and non-moving sources under one representation.

\paragraph{\revision{Training objectives.}}
\revision{For the final trajectory-training phase, the per-example dynamic loss is}
{\color{revisioncolor}
\begin{equation}
\begin{aligned}
\mathcal{L}_{\mathrm{dyn}} ={}&
\mathcal{L}_{\mathrm{sem}} + \mathcal{L}_{\mathrm{dir}}
+ 0.3\mathcal{L}_{\mathrm{dist}} \\
&+ 0.05\mathcal{L}_{\Delta\mathrm{dist}}
+ 0.05\mathcal{L}_{\Delta\mathrm{dir}}
+ 0.1\mathcal{L}_{\mathrm{act}}.
\end{aligned}
\end{equation}
}
\revision{Here, $\mathcal{L}_{\mathrm{sem}}$ and $\mathcal{L}_{\mathrm{act}}$ are binary cross-entropy losses for clip-level AudioSet labels and per-bin source activity, respectively.
$\mathcal{L}_{\mathrm{dir}}$ is the cosine loss between predicted and target 3D direction vectors, and $\mathcal{L}_{\mathrm{dist}}$ is the Smooth L1 loss on log-distance.
$\mathcal{L}_{\Delta\mathrm{dir}}$ and $\mathcal{L}_{\Delta\mathrm{dist}}$ apply Smooth L1 loss between predicted and target first-order differences of direction and log-distance across adjacent time bins, respectively.
The full objective additionally includes $20\mathcal{L}_{\mathrm{KD}}$, which distills semantic logits from the frozen Stage-1 teacher with temperature 1, and $0.05\mathcal{L}_{\mathrm{replay}}$, which applies the trajectory losses to replayed static clips represented as constant trajectories.}

\subsection{Spatio-Temporal Audio Tokens for the LLM}
\label{app:llm-connector}

During QA finetuning, the encoder prediction heads are removed and the frozen encoder is used as a representation extractor.
The language model receives the semantic global representation and the 40 temporally ordered trajectory representations.
The static direction and distance global representations are used for encoder training, but are not passed to the LLM.

The text prompt contains a single audio placeholder.
At the embedding level, this placeholder is replaced by the 41-representation audio block.
A trainable connector maps each audio representation to the LLM hidden size, and the projected block is inserted into the prompt embedding sequence.
\revision{The two-layer connector maps each 768-dimensional encoder representation to 4,096 dimensions and then applies a second 4,096-dimensional projection, with GELU activation and no dropout.}
The connector is trained jointly with LoRA adapters on the QA curriculum, while the audio encoder and the base LLM remain frozen.

\subsection{Baseline Implementations}
\label{app:baseline-implementations}

\paragraph{Encoder baselines.}
We compare \encodername{} against three classes of encoder baselines: a non-learned localization method, a static spatial-audio encoder, and a localization-focused dynamic encoder.

\textit{Intensity-based DOA estimator.}
This baseline is a non-learned method with no trainable parameters.
It estimates source direction from \foa{} active-intensity cues computed from the rendered \foa{} waveform.
Because it does not contain an event classifier, distance head, or language-facing representation, it is used only as a physics-based localization reference.

\textit{Spatial-AST-\foa{} temporal crop.}
We build a Spatial-AST encoder for static \foa{} input.
The architecture follows \cite{bat2024}, with the frontend modified to accept \foa{} features.
For dynamic evaluation, the moving-source clip is evaluated through temporal crops, and the static localization heads are applied to each crop.
This gives the baseline local temporal evidence while keeping the model architecture static.
It does not include the 40-bin temporal modeling block, dynamic trajectory heads, or moving-source trajectory supervision used by \encodername{}.

\textit{PSELDNets-mACCDOA with semantic and distance heads.}
We adapt the PSELDNets mACCDOA architecture as a localization-focused dynamic baseline.
The original model is designed around frame-wise sound event localization with a smaller event taxonomy, so we extend it with a 355-class AudioSet semantic head and an additional distance head.
For the encoder evaluation, we use ground-truth-class track selection: localization is evaluated on the output associated with the ground-truth event class.
This focuses the comparison on localization and representation quality rather than on an additional class-track selection problem.

\paragraph{Audio-language baselines.}
We compare \modelname{} with several audio-language baselines that differ in the audio representation supplied to the language model.
\revision{The QA-trained baselines use \datasetname{} supervision, and all models are evaluated with the same answer parser.}

\textit{BAT.}
The BAT baseline~\cite{bat2024} is a representative prior spatial-audio LLM designed for static spatial sources.
We evaluate it under our QA protocol using the same QA pairs and scene/source metadata as \modelname{}, but render the audio in the binaural format expected by BAT.
We train it with the same QA supervision to compare against a prior static spatial-audio language modeling pipeline.

\textit{PSELDNets-mACCDOA + OLMo2.}
This baseline connects the adapted PSELDNets-mACCDOA~\cite{hu2025pseldnets} encoder to the same OLMo2 language model~\cite{olmo20242}.
\revision{We extract the shared final HTSAT feature map rather than final SELD logits, average it over frequency and linearly resample it to 40 temporal tokens, and append its global average as one clip token.}
We then train the same connector and LoRA adapters used by \modelname{}.
This tests whether localization-focused dynamic features are sufficient for source-time-space QA when attached to the same LLM training pipeline.

\textit{Spatial-AST-\foa{} + OLMo2.}
This baseline connects the frozen static \foa{} encoder to OLMo2.
To match the audio-token count of \modelname{}, we reshape the post-transformer patch sequence into a time-frequency grid, average over frequency, resample the resulting temporal sequence to 40 positions, and concatenate the semantic global representation.
Thus, the LLM receives 41 audio representations, as in \modelname{}.
The key difference is that these temporal representations come from a static encoder and are not trained with dynamic trajectory heads or moving-source supervision.

\revision{\textit{Zero-shot audio-language models.}
We evaluate Qwen2-Audio~\cite{chu2024qwen2}, Audio Flamingo 3~\citep{goel2025audioflamingo3}, MOSS-Audio-8B~\citep{yang2026mossaudio}, and DIFFA-2~\citep{zhou2026diffa2} from their released instruction checkpoints without task-specific finetuning.
Qwen2-Audio receives SoundSpaces-rendered left and right binaural channels as separate audio inputs; the other three receive the omnidirectional \foa{} $W$ channel as 16-kHz mono.
All use the same QA questions, valid-answer format, and answer normalization.
The three additional models generate at most 24 tokens; Audio Flamingo 3 and MOSS-Audio-8B use greedy decoding, while DIFFA-2 uses 16 diffusion steps.}

\revision{\textit{Spatial-Omni.}
We evaluate the released SO-7B and SO-Encoder checkpoints~\citep{zhu2026spatialomni} without \datasetname{} adaptation.
Its native preprocessing receives 10-second, 16-kHz four-channel \foa{} audio, and generation uses the same exact-answer prompts and scoring protocol.}
\subsection{Training Configuration}
\label{app:training-hyperparameters}

\revision{The trajectory-learning stage of \encodername{} (Stage 3) proceeds in two phases.
We first train the encoder on 20K single-source moving clips for 50 epochs with a learning rate of $1\times10^{-5}$.
We then continue training on 100K reflected moving-source clips for 20 epochs with a learning rate of $3\times10^{-5}$.
In this final phase, we apply semantic supervision and distillation from the frozen Stage-1 semantic teacher, while replaying static examples as constant trajectories.
Both phases use AdamW with weight decay $5\times10^{-4}$, and all dynamic examples use 40 time bins per 10-second clip.}
    
For audio-language training, we use OLMo2-7B-Instruct as the base language model.
The audio encoder and the base LLM weights are frozen.
We train a two-layer MLP connector together with LoRA adapters.
\revision{The Type-A, Type-B, and Type-C curriculum stages are each trained for two epochs.}
The LoRA adapters use rank 16, $\alpha=32$, and dropout 0.05.
\revision{Type A uses connector and LoRA learning rates of $2\times10^{-4}$ and $10^{-4}$, while Types B and C use $10^{-4}$ and $5\times10^{-5}$, respectively.
The OLMo2-based QA-trained baselines use the same stage-wise connector and LoRA schedule.}

\section{Evaluation Details}
\label{app:evaluation-details}

All models are evaluated with the same answer parser and the same family-level aggregation.

\paragraph{Family-level aggregation.}
\revision{Detailed scores are computed separately for each question family.
Overall is accuracy over all questions in the corresponding evaluation pool after answer normalization.}
For Type B, source-to-attribute questions contribute to the corresponding attribute column, such as spatial location, spatial change, or movement status.
In contrast, attribute-to-source questions are evaluated under source grounding, since the model must select which source satisfies a queried attribute.
For Type C, Avg. is the unweighted average over the reported compositional reasoning families.

\paragraph{Event metrics.}
Event-detection questions are evaluated as multi-label \map{}.
For \datasetname{}, the positive labels are the accepted AudioSet labels attached to the queried source or source mixture; for two-source clips, the positive set is the union of both sources' retained labels.
\revision{We lowercase each generated answer, remove wrapper text and surrounding punctuation, and retain its first output line.
We then represent the answer and every canonical event label by TF-IDF-weighted token counts and use their cosine similarity as the score for that label.
Because this scoring is token based rather than exact-set matching, multi-label answers may use commas, semicolons, slashes, or conjunctions without requiring a fixed separator.
Comma-separated AudioSet display-name synonyms are treated as aliases when mapping target metadata to canonical class indices; generated variants are scored by their token similarity to the canonical label rather than by a separate exact-match synonym table.}
\revision{We compute AP separately for every class with at least one positive example and report the macro-average over these classes.}
Event-verification questions ask whether a specific event is present and are scored separately as binary yes/no accuracy.
For STARSS23, positives are the annotated STARSS23 event classes, so STARSS23 \map{} is computed within its 13-class taxonomy and is not directly comparable to the 355-class AudioSet \map{}.

\paragraph{\revision{Direction aggregation.}}
\revision{In the QA tables, DoA is the arithmetic mean of the azimuth- and elevation-question accuracies, while $\Delta$DoA averages the corresponding beginning-to-end azimuth- and elevation-change accuracies.}

\paragraph{Closed-answer parsing and random baselines.}
All non-event-detection questions use finite answer spaces.
We lowercase model outputs, strip surrounding punctuation, and map the result to the answer vocabulary of the corresponding family; unmapped outputs are counted as incorrect.
Random baselines use the same family aggregation.
For closed-answer questions, the random predictor samples uniformly from the valid answer space of each question.
\revision{For event \map{}, random chance is the macro-average class prevalence over classes with positives in the corresponding evaluation split, corresponding to chance-level AP under random scoring.}

\section{\revision{Curriculum versus Joint Training}}
\label{app:additional-results}
\label{app:curriculum-vs-joint}

\begin{table}[t]
\centering
\scriptsize
\color{revisioncolor}
\setlength{\tabcolsep}{3.0pt}
\renewcommand{\arraystretch}{1.08}
\resizebox{\columnwidth}{!}{%
\begin{tabular}{c l c c c}
\toprule
Stage & Training data & Type-A eval & Type-B eval & Type-C eval \\
\midrule
I & A & 76.5 & 32.7 & 13.7 \\
II & A+B & \textbf{76.8} & \textbf{53.0} & 10.2 \\
III & A+B+C & 76.3 & 52.6 & \textbf{67.5} \\
\bottomrule
\end{tabular}%
}
\caption{\textbf{Stage-wise QA curriculum evaluation.}
The model after each stage is evaluated on the same held-out pools. Type-A/B values report accuracy over all questions in their respective pools after answer normalization, while Type C is the unweighted average over its three compositional families.}
\label{tab:curriculum-progression}
\end{table}

\revision{Table~\ref{tab:curriculum-progression} shows the progression across the QA curriculum: Stage I establishes single-source perception, Stage II substantially improves two-source grounding, and Stage III enables compositional reasoning while largely preserving Type-A/B performance.}

\begin{table*}[t]
\centering
\scriptsize
\setlength{\tabcolsep}{1.7pt}
\renewcommand{\arraystretch}{1.05}
\resizebox{\linewidth}{!}{%
\begin{tabular}{l c c c c c c c c c c c c c c c c c c c c c c}
\toprule
\multirow{3}{*}{Training strategy} &
\multirow{3}{*}{Epoch} &
\multicolumn{7}{c}{\textbf{A: Single-source perception}} &
\multicolumn{10}{c}{\textbf{B: Two-source perception and grounding}} &
\multicolumn{4}{c}{\textbf{C: Compositional reasoning}} \\
\cmidrule(lr){3-9}\cmidrule(lr){10-19}\cmidrule(lr){20-23}
& & \multicolumn{2}{c}{\textbf{Semantic}} &
\multicolumn{2}{c}{\textbf{Spatial loc.}} &
\multicolumn{2}{c}{\textbf{Spatial chg.}} &
\textbf{Move}
& \multicolumn{2}{c}{\textbf{Semantic}} &
\multicolumn{2}{c}{\textbf{Spatial loc.}} &
\multicolumn{2}{c}{\textbf{Spatial chg.}} &
\textbf{Move} &
\multicolumn{3}{c}{\textbf{Source grounding}} &
Temp. rel. & Move-spat. & Traj. rel. & Avg. \\
\cmidrule(lr){3-4}\cmidrule(lr){5-6}\cmidrule(lr){7-8}\cmidrule(lr){9-9}
\cmidrule(lr){10-11}\cmidrule(lr){12-13}\cmidrule(lr){14-15}\cmidrule(lr){16-16}\cmidrule(lr){17-19}
\cmidrule(lr){20-23}
& & \map{} & Y/N & DoA & Dist. & $\Delta$DoA & $\Delta$Dist. & Acc.
& \map{} & Y/N & Src. DoA & Src. Dist. & Src. $\Delta$DoA & Src. $\Delta$Dist. & Src. Move
& Loc. & Chg. & Move
& Acc. & Acc. & Acc. & Avg. \\
\midrule
Curriculum A+B+C & 6 (2+2+2)
& \textbf{27.6} & \textbf{93.7} & \textbf{81.6} & \textbf{51.5} & \textbf{91.1} & \textbf{87.8} & \textbf{99.8}
& \textbf{14.3} & \textbf{83.8} & \textbf{54.3} & \textbf{32.8} & \textbf{71.6} & \textbf{70.0} & 66.2
& \textbf{49.7} & \textbf{47.4} & \textbf{59.3}
& \textbf{86.0} & 55.8 & 60.6 & 67.5 \\
Joint A+B+C & 6
& 25.0 & 93.5 & 79.8 & 48.7 & 87.7 & 85.4 & 99.7
& 12.9 & 82.5 & 52.0 & 30.6 & 69.5 & 66.8 & \textbf{67.7}
& 48.2 & 47.4 & 59.0
& 85.0 & \textbf{57.6} & \textbf{61.0} & \textbf{67.9} \\
\bottomrule
\end{tabular}%
}
\caption{\textbf{Curriculum versus joint QA training.}
All values are on a 0--100 scale.
The curriculum checkpoint follows the staged schedule used in the main experiments and reports the same Type-C checkpoint as Table~\ref{tab:typec-results}.
The joint checkpoint is trained from scratch on the combined A+B+C pool.
Type-C Avg. is computed over the three reported Type-C families.}
\label{tab:curriculum-joint}
\end{table*}

Table~\ref{tab:curriculum-joint} compares the staged QA curriculum with a joint A+B+C training run under the same total number of QA-training epochs.
The joint run trains on the combined QA pool from the beginning, whereas the curriculum introduces single-source perception, two-source grounding, and compositional reasoning sequentially.
This ablation tests whether staged training better preserves the perception and grounding skills learned in Types A and B while adding Type-C compositional reasoning.

Joint training reaches comparable Type-C accuracy, and is slightly higher on the Type-C average.
However, under the same epoch budget, the staged curriculum consistently retains stronger Type-A and Type-B semantic, spatial-location, and spatial-change performance.
\revision{This indicates that staged training better preserves earlier source-state perception and grounding while adding compositional reasoning.}
We therefore use the staged curriculum for the main model and report joint training as an ablation.

\section{Real-Data Transfer}
\label{app:real-data-transfer}

\revision{We evaluate real-world transfer in two complementary settings: transfer and lightweight adaptation on STARSS23-derived QA~\citep{shimada2023starss23}, and zero-shot evaluation on STAR-Bench~\citep{liu2025starbench}.}

\subsection{STARSS23 Transfer QA Construction}
\label{app:starss23-transfer}

We construct a transfer benchmark from the original STARSS23~\cite{shimada2023starss23} frame-level annotations.
Each 10-second clip is paired with its \foa{} waveform and annotations for event class, source track, azimuth, elevation, and distance.
We define a physical source as a \texttt{(class, track)} pair within a clip.
For source-grounded questions, a class name is used as a referent only when that class appears in a single track; repeated class instances are treated as ambiguous for source selection but remain valid for event detection and verification.

\revision{As summarized in Table~\ref{tab:starss23-transfer-qa}, the STARSS23 transfer set contains Type A, Type B, and Type C questions satisfying the stated validity criteria.}
Type A is generated from clips with exactly one annotated source and evaluates event semantics, event verification, movement status, spatial location, and spatial change.
Type B is generated from clips with two or more annotated sources and evaluates the same perceptual attributes together with source grounding.
\revision{For Type C, we retain movement-conditioned spatial relation and cross-source trajectory relation, the two compositional families supported by the real annotations.
We omit temporal relation change because the same validity criteria yield insufficient balanced instances.}

\begin{table}[t]
\centering
\small
\begin{tabularx}{\columnwidth}{lrrX}
\toprule
\textbf{QA type} & \textbf{Train QA} & \textbf{Test QA} & \textbf{Question families} \\
\midrule
Type A & 4{,}003 & 1{,}775 & event, movement, location, change \\
Type B & 49{,}163 & 12{,}328 & event, movement, location, change, grounding \\
\revision{Type C} & \revision{1{,}075} & \revision{436} & \revision{movement--spatial and trajectory relations} \\
\midrule
Total & \revision{54{,}241} & \revision{14{,}539} & \revision{joint Type A/B/C} \\
\bottomrule
\end{tabularx}
\caption{\textbf{STARSS23-derived transfer QA split.}
\revision{Type C includes the two compositional families that remain reliable after filtering the real annotation distribution.}}
\label{tab:starss23-transfer-qa}
\end{table}

Temporal anchors follow the same linguistic convention as \datasetname{}: questions use \textit{beginning}, \textit{middle}, and \textit{end} without exposing numeric timestamps.
For label generation, we use the nearest active annotation around the queried anchor and instantiate a question only when the relevant source is active.
Movement labels are computed from the first and last active annotations of a source.
A source is considered moving if its azimuth change is at least 10 degrees, elevation change is at least 10 degrees, distance change is at least 0.5 m, or relative distance change is at least 20\%.

\subsection{STARSS23 Transfer and Adaptation}
\label{app:starss23-results}
\revision{We aim to measure whether the synthetic source-time-space representation transfers to real recordings and how different uses of real QA affect transfer.
\textit{Zero-shot} directly evaluates the \modelname{} checkpoint trained on \datasetname{} without STARSS23 supervision.
\textit{Adapted} fine-tunes only the connector and LoRA adapters for two epochs on the corresponding STARSS23-derived QA training pools after the synthetic curriculum, keeping \encodername{} frozen.
\textit{Mixture} instead adds STARSS23-derived QA during each curriculum stage, with the number of real questions set to approximately 5\% of that stage's synthetic QA count.
This gives 10{,}822, 25{,}822, and 41{,}422 real questions in Stages 1, 2, and 3, respectively.
Using seed 0, we sample by question family: Stages 1--2 use replacement when a family does not meet its quota, while Stage 3 samples without replacement and redistributes unused family quotas across the remaining pool.}

\begin{table*}[t]
\centering
\scriptsize
\setlength{\tabcolsep}{1.7pt}
\renewcommand{\arraystretch}{1.08}
\resizebox{\linewidth}{!}{%
\begin{tabular}{l l c c c c c c c @{\hspace{6pt}} c c c c c c c c c}
\toprule
\multirow{3}{*}{Setting} &
\multirow{3}{*}{Model} &
\multicolumn{7}{c}{\textbf{A: Single-source perception}} &
\multicolumn{9}{c}{\textbf{B: Multi-source perception and grounding}} \\
\cmidrule(lr){3-9}\cmidrule(lr){10-18}
& & \multicolumn{2}{c}{\textbf{Semantic}} & \multicolumn{2}{c}{\textbf{Spatial loc.}} & \multicolumn{2}{c}{\textbf{Spatial chg.}} & \textbf{Move}
& \multicolumn{2}{c}{\textbf{Semantic}} & \multicolumn{2}{c}{\textbf{Spatial loc.}} & \multicolumn{2}{c}{\textbf{Spatial chg.}} & \textbf{Move} & \multicolumn{2}{c}{\textbf{Source grd.}} \\
\cmidrule(lr){3-4}\cmidrule(lr){5-6}\cmidrule(lr){7-8}\cmidrule(lr){9-9}\cmidrule(lr){10-11}\cmidrule(lr){12-13}\cmidrule(lr){14-15}\cmidrule(lr){16-16}\cmidrule(lr){17-18}
& & \map{} & Y/N & DoA & Dist. & $\Delta$DoA & $\Delta$Dist. & Acc.
& \map{} & Y/N & DoA & Dist. & $\Delta$DoA & $\Delta$Dist. & Acc. & DoA & Dist. \\
\midrule
-- & Random chance & 12.50 & 50.00 & 22.92 & 12.50 & 50.00 & 50.00 & 50.00
& 23.22 & 50.00 & 22.92 & 12.50 & 50.00 & 50.00 & 18.25 & 17.74 & 17.68 \\
\addlinespace[1pt]
\midrule
Zero-shot & \modelname{} & 49.47 & 83.67 & 26.51 & 25.46 & 40.91 & 40.00 & 88.67
& 42.96 & 72.24 & 19.74 & 11.79 & 51.19 & 54.29 & 1.49 & 12.07 & 8.18 \\
\revision{Light adapt.} & \modelname{} + STARSS23 & 73.52 & 83.67 & 59.97 & 32.55 & 50.00 & 30.00 & 92.67
& 59.81 & 87.47 & 53.46 & 33.93 & 51.19 & 55.71 & 69.78 & 60.77 & 67.49 \\
\revision{Real mixture} & \revision{\modelname{} + 5\% STARSS23} & \revision{71.81} & \revision{94.00} & \revision{68.24} & \revision{37.53} & \revision{63.64} & \revision{30.00} & \revision{93.33}
& \revision{67.48} & \revision{88.67} & \revision{56.88} & \revision{32.91} & \revision{48.81} & \revision{52.86} & \revision{73.51} & \revision{57.89} & \revision{64.72} \\
\bottomrule
\end{tabular}%
}
\caption{\textbf{\revision{Detailed STARSS23 Type-A/B results.}}
All values are on a 0--100 scale and higher is better.
Semantic \map{} is computed over the 13 STARSS23 event classes.
DoA averages azimuth and elevation accuracy.
Source grd. evaluates Type-B attribute-conditioned source selection.
\revision{\textit{Real mixture} adds a family-stratified STARSS23 QA set amounting to approximately 5\% of the synthetic QA count at each curriculum stage; exact counts and sampling are described above.}}
\label{tab:starss23-transfer-results}
\end{table*}

\revision{Table~\ref{tab:starss23-transfer-results} provides a fine-grained Type-A/B breakdown complementing the overall and Type-C results in Table~\ref{tab:real-recording-results}(a).
The zero-shot model transfers event recognition, but source-time-space grounding remains difficult without real-data supervision.
Light adaptation substantially improves Type-B localization, movement, and source grounding.
The real-data mixture yields the strongest Type-A overall result and matches light adaptation on Type-B overall (59.1 vs. 59.4), with complementary strengths across individual metrics.
As shown in Table~\ref{tab:real-recording-results}(a), light adaptation remains stronger on Type-C compositional reasoning.}

\subsection{\revision{STAR-Bench Zero-Shot Evaluation}}
\label{app:starbench-results}

\begingroup
\color{revisioncolor}
We use the 123 STARSS23-derived items of STAR-Bench~\citep{liu2025starbench} whose released clips are at most 10 seconds: 101 single-source static-localization, 12 multi-source spatial-relation, and 10 dynamic trajectory-tracking items.
We align each released clip to its original STARSS23 recording by short-time envelope correlation and zero-pad the corresponding \foa{} segment to 10 seconds.
Each three-option question is evaluated under all three cyclic option orders, yielding 369 trials and reducing answer-position bias.
Models answer with the exact option text and use their native or closest compatible interface: \foa{} for \foa{}-capable models, separate binaural channels for Qwen2-Audio, and mono for Audio Flamingo 3.
All models are evaluated zero-shot, and no STAR-Bench data is used to train \modelname{}.

As reported in Table~\ref{tab:real-recording-results}(b), \modelname{} achieves the best static-localization, relation, trajectory, and overall results.
Together with STARSS23 transfer, this zero-shot result provides complementary evidence that the trajectory-aware representation generalizes beyond the synthetic rendering pipeline.
\endgroup

\section{Data Generation Details}
\label{app:data-details}

\subsection{\foa{} Scene and Dynamic Audio Rendering}
\label{app:rendering-details}

\paragraph{Listener and source placement.}
We sample listener and source anchors from precomputed navigable locations in each Matterport3D scene~\cite{chang2017matterport3d}.
Before RIR rendering, we enforce a same-room constraint on each listener--source pair and discard pairs that cross room boundaries, preventing them from entering either encoder training or QA generation.
For a moving source, the listener remains fixed and the source trajectory is constructed from valid source anchors in the same room.
For a static source, we use a constant trajectory by repeating the same source position across all time bins.

\paragraph{Time-varying rendering.}
SoundSpaces~\cite{chen2022soundspaces2} provides \foa{} RIRs for fixed source-listener configurations.
We therefore approximate a moving source as a sequence of short stationary renderings.
Each 10-second clip is divided into $N=40$ time bins, with $\Delta=10/N=0.25$ seconds.
Let $\tau_j=(j+\frac{1}{2})\Delta$ denote the center of bin $j$, and let $\mathbf{p}_k(\tau_j)$ be the 3D position of source $k$ at that time.
For each bin, SoundSpaces is queried for the \foa{} RIR $h_{k,m,j}$ from source $k$ to ambisonic channel $m$ at position $\mathbf{p}_k(\tau_j)$.

Given a monaural event waveform $x_k$, we render the $m$-th \foa{} channel of source $k$ by convolving $x_k$ with the per-bin RIRs and crossfading the rendered outputs:
\begin{equation}
    \tilde{x}_{k,m}(t)=
    \sum_{j=0}^{N-1} w_j(t)\,
    \bigl(h_{k,m,j} \ast x_k\bigr)(t),
\end{equation}
where $w_j(t)$ are 50 ms crossfade weights satisfying $\sum_j w_j(t)=1$ for all $t$.
\revision{This crossfaded-RIR strategy follows prior moving-source SELD rendering~\citep{politis2020dataset}.
The maximum radial source velocity in our trajectory design is $v_r=0.28$ m/s, yielding a relative Doppler shift of $|v_r|/c \approx 8.2{\times}10^{-4}$.
Thus, the omitted frequency shift is below 0.1\% in our motion regime and is unlikely to dominate direction/distance trajectory QA; high-speed pass-by events remain outside our scope.}
The rendered source signal is
$\tilde{\mathbf{x}}_k(t)=\{\tilde{x}_{k,m}(t)\}_m$ over all \foa{} channels.
For two-source scenes, source stems are rendered independently and summed channel-wise:
\begin{equation}
    \tilde{\mathbf{x}}^{\mathrm{scene}}(t)
    =
    \sum_{k=1}^{K} \tilde{\mathbf{x}}_k(t).
\end{equation}

\paragraph{Trajectory filtering.}
\revision{We retain render-valid trajectories with finite, nonsilent RIRs.
A source is dynamic when its active beginning and end anchors differ by at least $10^\circ$ in azimuth or elevation, or 0.5 m in distance; a static source is formed by freezing a randomly selected beginning, middle, or end state.
Two-source scenes use distinct source-audio, trajectory, and canonical event identifiers under the same scene, room, and listener, with mean and minimum inter-source angular separations of at least $35^\circ$ and $10^\circ$, respectively.}

\paragraph{Activity-aware QA labeling.}
We generate QA labels only when the queried source is audible at the requested time anchor.
To determine this, we compute a source-level activity mask before mixing sources.
This is important for two-source clips: at the same time anchor, one source may be active and valid for a question, while the other may be inactive and therefore skipped.
This prevents generating questions whose answers would depend on inaudible or near-silent source segments.

\paragraph{Metadata for QA generation.}
For each rendered clip, the manifest stores all information needed to generate and verify QA labels.
This includes the scene identifier, listener pose, source trajectories, source-to-listener directions and distances, source activity masks, AudioSet labels, and canonical source labels.
It also stores quantized spatial states at the beginning, middle, and end of the clip, which are used to generate time-anchored localization questions.
Finally, beginning-to-end spatial deltas are stored to determine whether spatial-change questions are valid.

\subsection{QA Generation Rules and Templates}
\label{app:qa-generation-details}

\begin{table}[t]
\centering
\small
\begin{tabular}{lll}
\toprule
\textbf{Azimuth label} & \textbf{Center angle} & \textbf{Angular sector} \\
\midrule
Front (F)        & $0^\circ$   & $[-22.5^\circ, 22.5^\circ)$ \\
Front-right (FR) & $45^\circ$  & $[22.5^\circ, 67.5^\circ)$ \\
Right (R)        & $90^\circ$  & $[67.5^\circ, 112.5^\circ)$ \\
Back-right (BR)  & $135^\circ$ & $[112.5^\circ, 157.5^\circ)$ \\
Back (B)         & $180^\circ$ & $[157.5^\circ, 202.5^\circ)$ \\
Back-left (BL)   & $225^\circ$ & $[202.5^\circ, 247.5^\circ)$ \\
Left (L)         & $270^\circ$ & $[247.5^\circ, 292.5^\circ)$ \\
Front-left (FL)  & $315^\circ$ & $[292.5^\circ, 337.5^\circ)$ \\
\bottomrule
\end{tabular}
\caption{\textbf{Azimuth-sector quantization.}
Azimuth is quantized into eight listener-centered 45-degree sectors. Angles are measured clockwise on the horizontal plane, with front at $0^\circ$.}
\label{tab:azimuth_quantization}
\end{table}

\paragraph{Answer quantization.}
We quantize continuous source states into listener-centered azimuth sectors, elevation categories, distance bins, and spatial-change labels.
For azimuth, we use eight listener-centered 45-degree sectors, as summarized in Table~\ref{tab:azimuth_quantization}.
Elevation labels are defined as \textit{above}, \textit{level}, and \textit{below}, with a $\pm10^\circ$ band for \textit{level}.
\revision{Distance labels use ten 0.5 m bins, from 0.5--1.0 m to 5.0--5.5 m.}
For spatial-change questions, we instantiate a question only when the beginning-to-end change exceeds the corresponding threshold.
Questions below these thresholds are skipped to avoid noisy or perceptually ambiguous labels.

\paragraph{Event labels and source referents.}
AudioSet clips can contain multiple valid labels.
For event detection, the accepted answer set is the full retained AudioSet label set; for two-source event detection, the accepted set is the union of the two sources' label sets.
For source-specific questions, however, each source also has one canonical event label used as the linguistic referent.
This keeps questions such as ``the clapping'' well-defined while still evaluating event recognition with multi-label targets.

\paragraph{Source-grounding validity.}
Two-source source-grounding questions use a label-choice answer space: the two canonical source labels, \textit{both}, and \textit{neither}.
The generator instantiates a source-grounding question only when the queried attribute is valid for the relevant source state and the answer is well-defined under this four-option space.
This is especially important for spatial attributes, since both sources may occupy the same sector or neither source may match the queried sector.

\paragraph{Type-C validity.}
Type C is generated from two-source clips after all primitive labels have been computed.
Temporal relation-change questions require the compared relation to be valid at both the beginning and end anchors.
Movement-conditioned spatial relations require a valid movement-status referent, such as a moving source and a stationary source, before the spatial comparison is formed.
Cross-source trajectory relations require both trajectories to have valid beginning-to-end spatial-change labels.

\subsection{\datasetname{} Statistics}
\label{app:dataset-statistics}
\label{app:dataset_statistics}

\revision{\paragraph{Dataset splits.}
We partition the eligible dynamic-source pool by base trajectory identifier into 80\%/10\%/10\% train/validation/test pools before constructing single- and two-source clips.
Balanced sampling then produces 40,000/4,000/10,000 clips: half single-source, equally split between static and dynamic, and half two-source, equally split among SS, SD, DS, and DD motion cases; SD and DS constitute the mixed static--dynamic configuration reported in the main text.
No source-audio or base-trajectory identifier is shared across splits.}

Table~\ref{tab:appendix_dataset_statistics} summarizes the number of generated QA pairs for each task family across the training, validation, and test splits.
The table separates single-source and two-source QA pairs, where Type A covers single-source perception QA, Type B covers two-source perception and grounding QA, and Type C covers compositional source-time-space reasoning QA.
The primitive QA families are grouped into event semantics, movement status, spatial location, spatial change, and source grounding.
Type C further introduces three compositional reasoning families: temporal relation change, movement-conditioned spatial relation, and cross-source trajectory relation.

\begin{figure*}[t]
    \centering
    \includegraphics[width=1\linewidth]{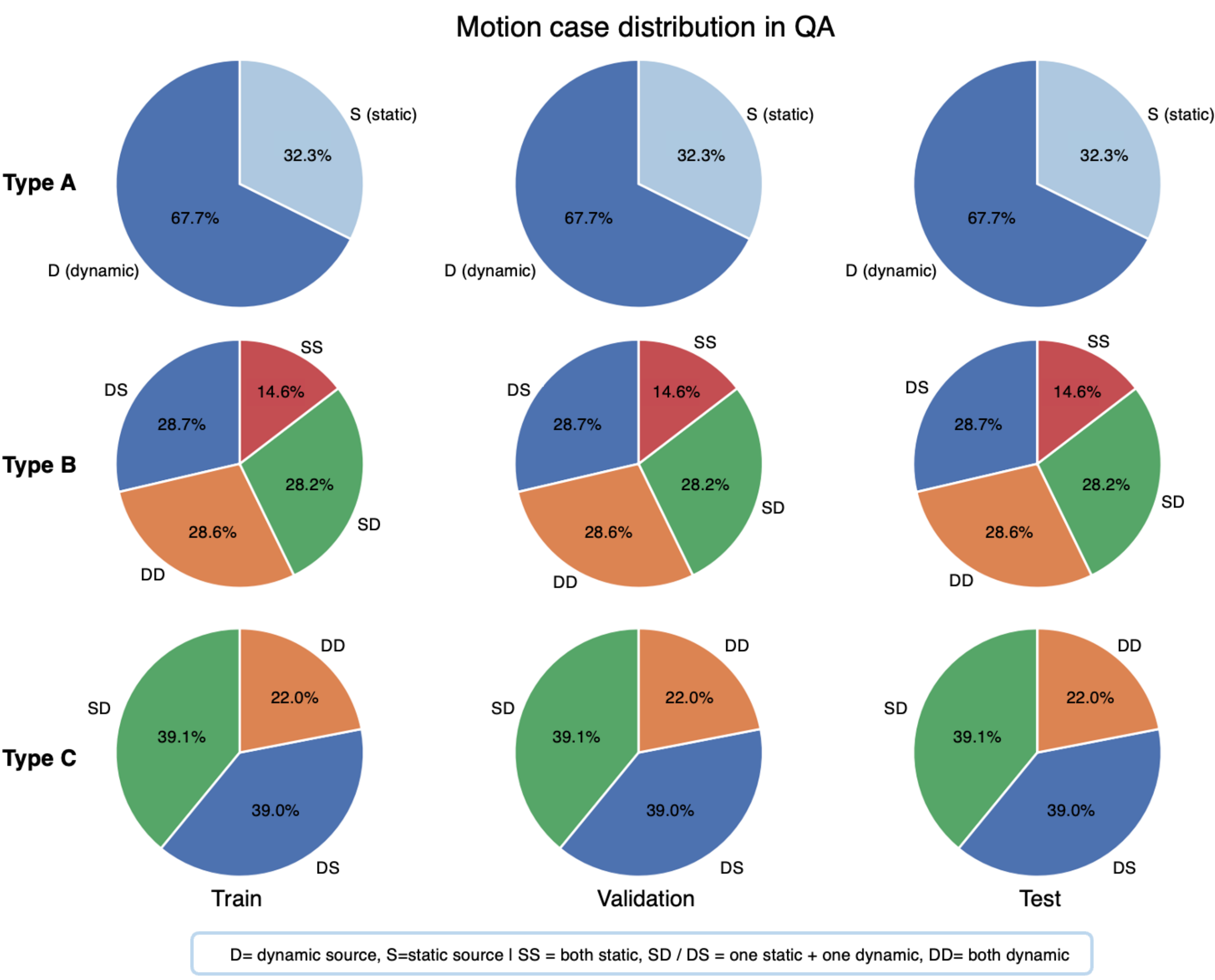}
    \caption{\textbf{Motion-case statistics.} We show distributions of motion cases for Type A, Type B, and Type C QA examples across the training, validation, and test sets.}
    \label{fig:stat_motion}
\end{figure*}

\begin{figure*}[t]
    \centering
    \includegraphics[width=1\linewidth]{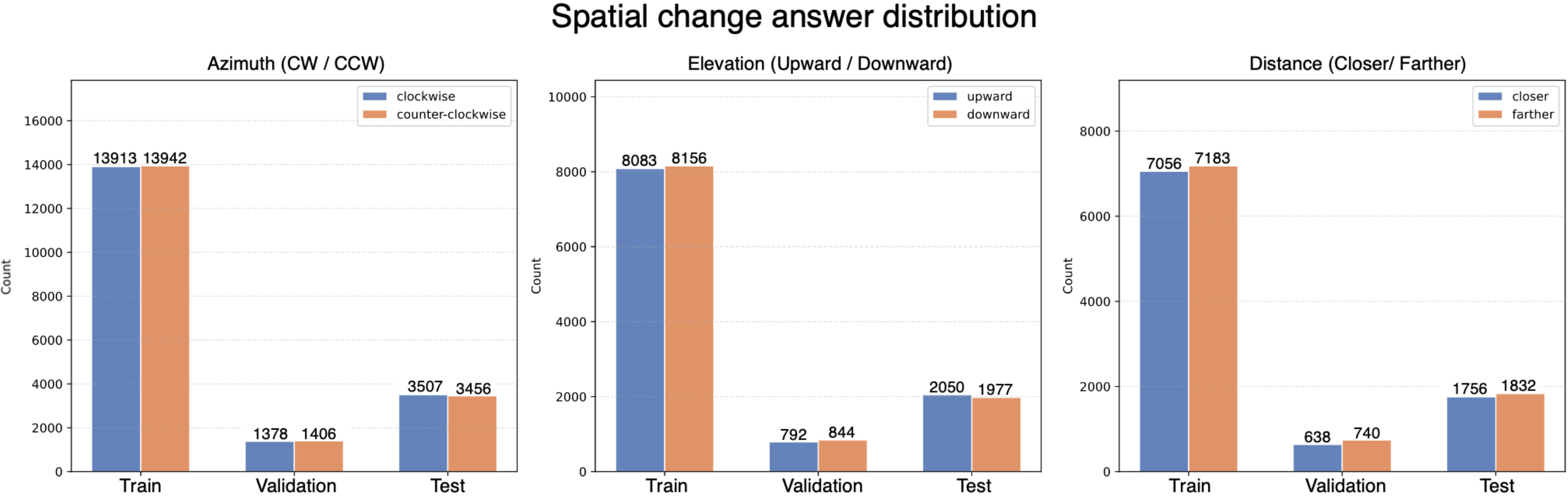}
    \caption{\textbf{Spatial-change annotation statistics.} We show distributions of spatial-change annotations along azimuth, elevation, and distance axes across the training, validation, and test sets.}
    \label{fig:stat_change}
\end{figure*}

\begin{figure*}[t]
    \centering
    \includegraphics[width=1\linewidth]{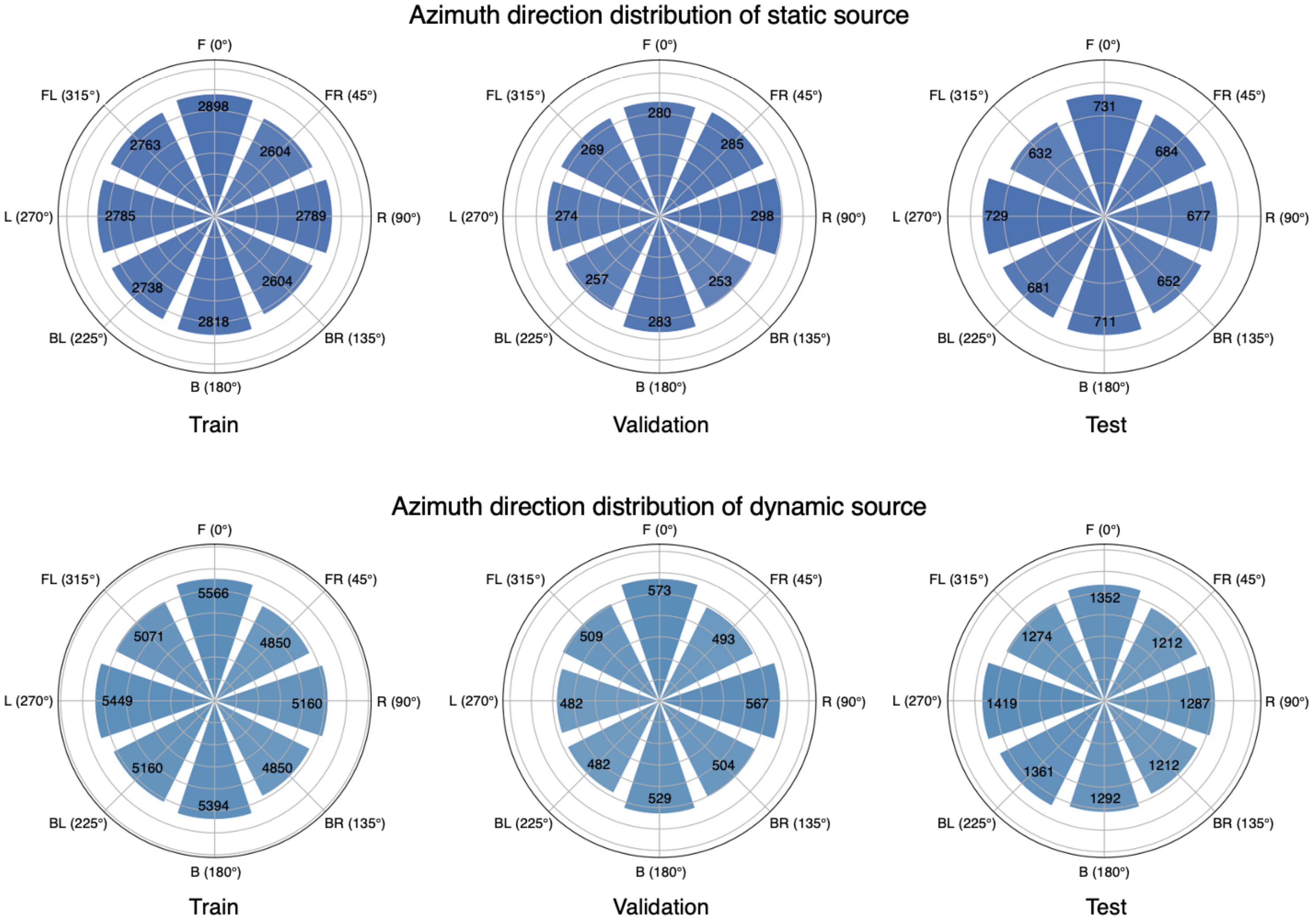}
    \caption{\textbf{Azimuth annotation statistics.} We show polar histograms of azimuth-direction annotations for static and dynamic sources across the training, validation, and test sets.}
    \label{fig:stat_azimuth}
\end{figure*}

\begin{figure*}[t]
    \centering
    \includegraphics[width=1\linewidth]{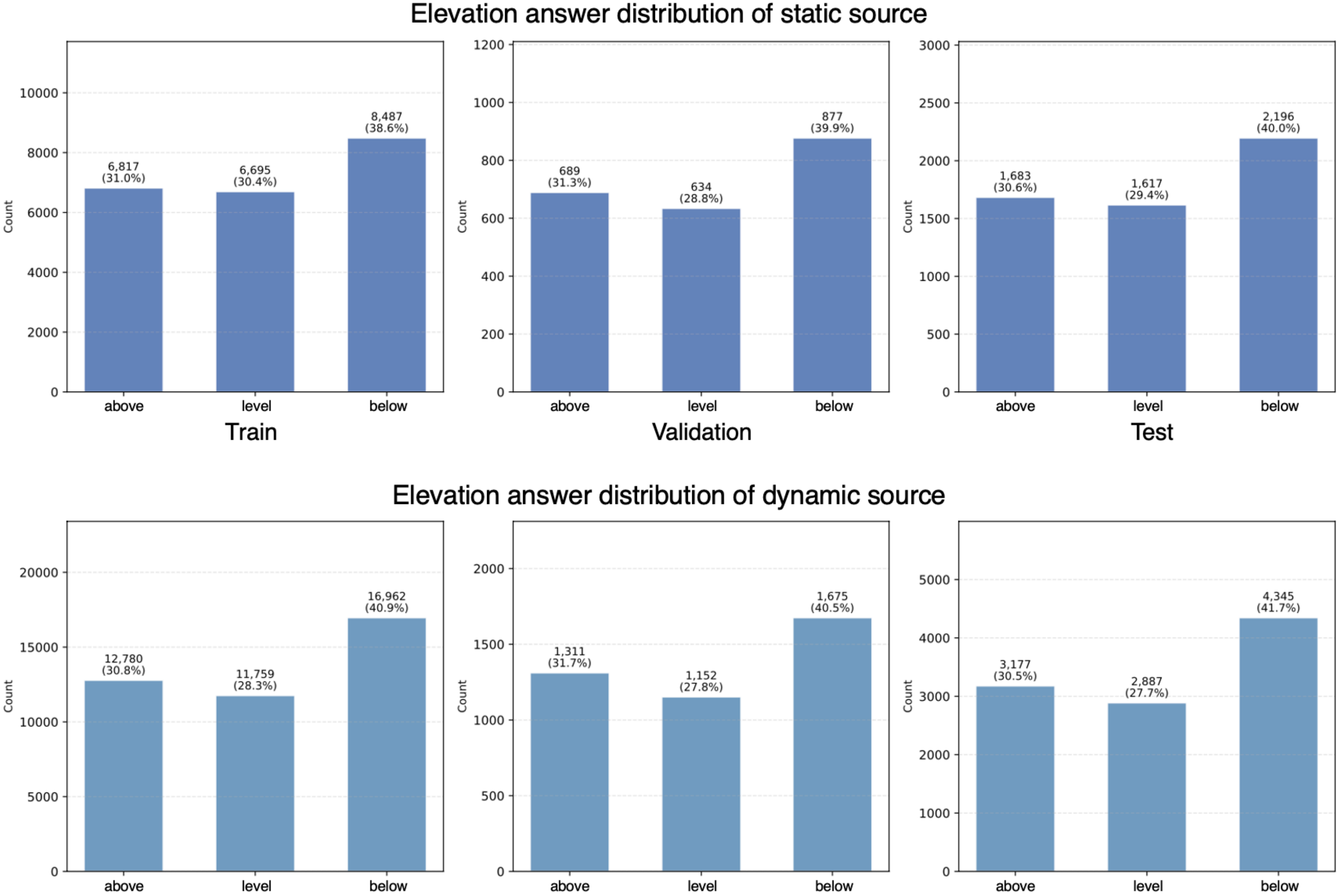}
    \caption{\textbf{Elevation annotation statistics.} We show distributions of elevation annotations for static and dynamic sources across the training, validation, and test sets.}
    \label{fig:stat_elevation}
\end{figure*}

\begin{figure*}[t]
    \centering
    \includegraphics[width=1\linewidth]{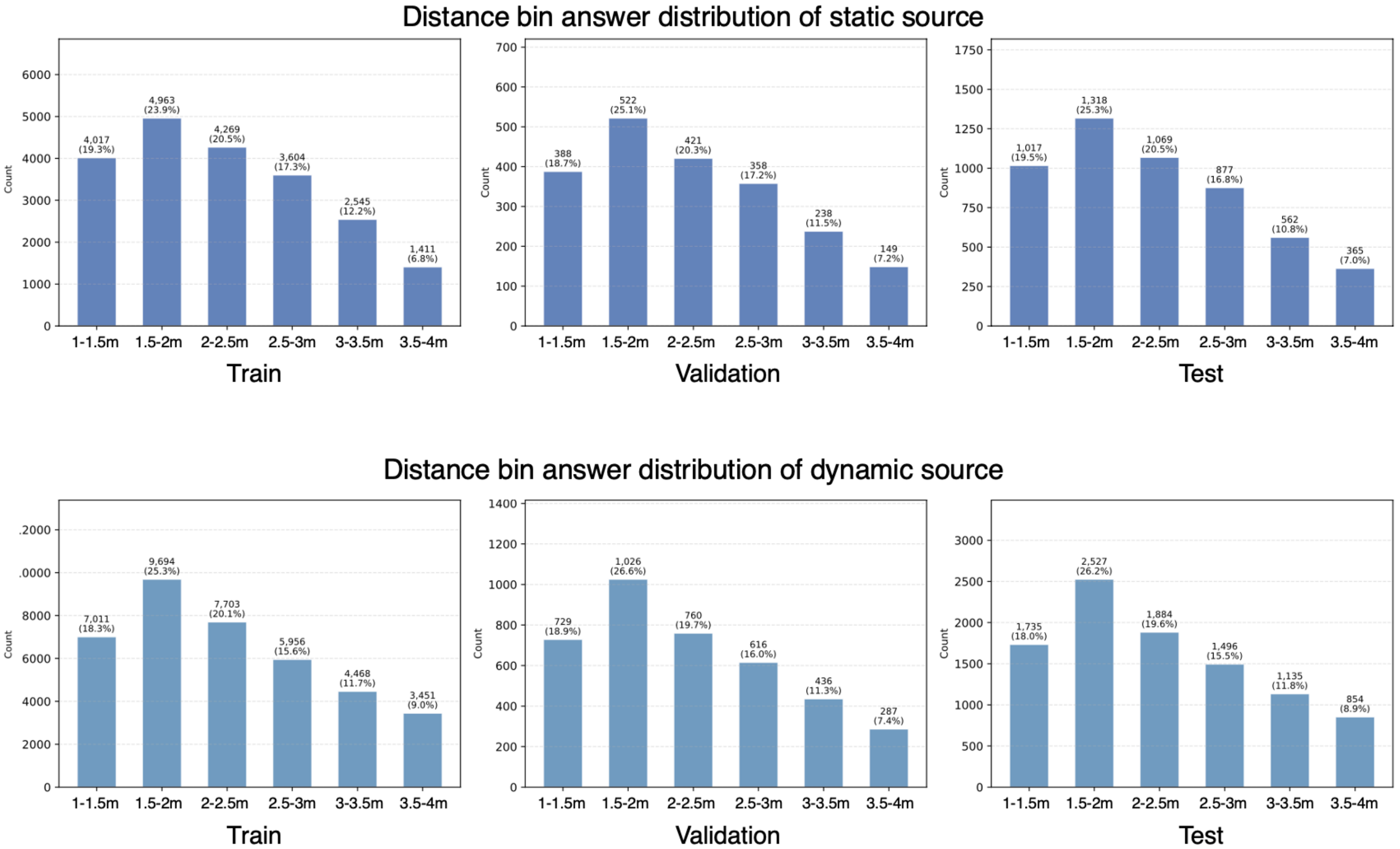}
    \caption{\textbf{Distance annotation statistics.} We show histograms of distance-bin annotations for static and dynamic sources across the training, validation, and test sets.}
    \label{fig:stat_dist}
\end{figure*}

Figures~\ref{fig:stat_motion}--\ref{fig:stat_dist} provide additional annotation statistics.
Figure~\ref{fig:stat_motion} shows the distribution of motion cases for Type A, Type B, and Type C across splits, highlighting the balance between static and dynamic source configurations.
Figure~\ref{fig:stat_change} visualizes the answer distributions for spatial-change questions along azimuth, elevation, and distance axes.
Figures~\ref{fig:stat_azimuth}, \ref{fig:stat_elevation}, and \ref{fig:stat_dist} show the distributions of azimuth, elevation, and distance annotations for static and dynamic sources.

\begin{table*}[t]
\centering
\scriptsize
\caption{\textbf{Dataset statistics by task split.}
Each entry reports the number of QA pairs.
Single and Two denote the number of single-source and two-source QA pairs, respectively.
Type A corresponds to single-source perception QA, Type B to two-source perception and grounding QA, and Type C to compositional source-time-space reasoning QA.
Primitive QA families are grouped into five high-level tasks, and compositional reasoning families into three relation tasks.}
\label{tab:appendix_dataset_statistics}
\resizebox{\linewidth}{!}{%
\begin{tabular}{@{\,}l@{\quad}l@{\quad}r@{\quad}r@{\quad}r@{\quad}|r@{\quad}r@{\quad}r@{\quad}|r@{\quad}r@{\quad}r@{\,}}
\toprule
\multirow{2}{*}{\textbf{Task family}}
& \multirow{2}{*}{\textbf{Task}}
& \multicolumn{3}{c|}{\textbf{Train}}
& \multicolumn{3}{c|}{\textbf{Validation}}
& \multicolumn{3}{c}{\textbf{Test}} \\
\cmidrule(lr){3-5}\cmidrule(lr){6-8}\cmidrule(lr){9-11}
& & \textbf{Single} & \textbf{Two} & \textbf{Total}
  & \textbf{Single} & \textbf{Two} & \textbf{Total}
  & \textbf{Single} & \textbf{Two} & \textbf{Total} \\
\midrule

\multirow{4}{*}{\makecell[l]{\textbf{Type A:} Single-source\\perception QA}}
& Event semantics
& 59,988 & 0 & 59,988
& 6,000 & 0 & 6,000
& 14,988 & 0 & 14,988 \\

& Movement status
& 19,996 & 0 & 19,996
& 2,000 & 0 & 2,000
& 4,996 & 0 & 4,996 \\

& Spatial location
& 116,100 & 0 & 116,100
& 11,574 & 0 & 11,574
& 29,115 & 0 & 29,115 \\

& Spatial change
& 20,353 & 0 & 20,353
& 2,034 & 0 & 2,034
& 5,082 & 0 & 5,082 \\

\midrule
\multirow{5}{*}{\makecell[l]{\textbf{Type B:} Two-source \\perception and \\ grounding QA}}
& Event semantics
& 0 & 50,000 & 50,000
& 0 & 5,000 & 5,000
& 0 & 12,500 & 12,500 \\

& Movement status
& 0 & 20,000 & 20,000
& 0 & 2,000 & 2,000
& 0 & 5,000 & 5,000 \\

& Spatial location
& 0 & 74,400 & 74,400
& 0 & 7,440 & 7,440
& 0 & 18,600 & 18,600 \\

& Spatial change
& 0 & 37,980 & 37,980
& 0 & 3,764 & 3,764
& 0 & 9,496 & 9,496 \\

& Source grounding
& 0 & 117,620 & 117,620
& 0 & 11,762 & 11,762
& 0 & 29,378 & 29,378 \\

\midrule
\multirow{3}{*}{\makecell[l]{\textbf{Type C:} Compositional\\source-time-space\\reasoning QA}}
& Temporal relation change
& 0 & 78,000 & 78,000
& 0 & 7,800 & 7,800
& 0 & 19,500 & 19,500 \\

& Movement-conditioned spatial relation
& 0 & 78,000 & 78,000
& 0 & 7,800 & 7,800
& 0 & 19,500 & 19,500 \\

& Cross-source trajectory relation
& 0 & 61,965 & 61,965
& 0 & 6,291 & 6,291
& 0 & 15,679 & 15,679 \\

\midrule
\textbf{Total}
& --
& \textbf{216,437} & \textbf{517,965} & \textbf{734,402}
& \textbf{21,608} & \textbf{51,857} & \textbf{73,465}
& \textbf{54,181} & \textbf{129,653} & \textbf{183,834} \\
\bottomrule
\end{tabular}%
}
\end{table*}

\end{document}